\documentclass[aps,twocolumn,groupedaddress]{revtex4}
\usepackage{epsfig}
\usepackage{graphicx}
\usepackage[T1]{fontenc}
\usepackage{ae}
\usepackage{color}
\usepackage[latin1]{inputenc}
\usepackage{amssymb,amsbsy,amsmath}
\usepackage{bbm}

\newtheorem{lemma}{Lemma}

\newtheorem{prop}{Proposition}

\newtheorem{theorem}{Theorem}

\begin{document}



\title[Theory of ground states III]
{Theory of ground states for classical Heisenberg spin
systems III}

\author{Heinz-J\"urgen Schmidt$^1$
\footnote[1]{Correspondence should be addressed to hschmidt@uos.de}
}
\address{$^1$Universit\"at Osnabr\"uck, Fachbereich Physik,
Barbarastr. 7, D - 49069 Osnabr\"uck, Germany}


\begin{abstract}
We extend the theory of ground states of classical Heisenberg spin systems published previously by a closer
investigation of the convex Gram set of $N$ spins.
This is relevant for the present purpose since the set of ground states of a given spin system corresponds to a face of the
Gram set. The investigation of the Gram set is facilitated by the determination of its symmetry group.
The case of the general spin triangle is completely solved and illustrated by a couple of examples.
\end{abstract}

\maketitle

\section{Introduction}\label{sec:I}
This is the third of a series of papers devoted to the theory of ground states of finite classical Heisenberg
spin systems. The general motivation of such a theory can be found in \cite{S17} and need
not be repeated here. Extended examples are contained in \cite{S17b}. In section \ref{sec:DLV} we will recapitulate the central definitions and results of
the ``Lagrange variety approach" presented in \cite{S17}. The system of $N$ spins will be described by a symmetric
$N\times N$-matrix ${\mathbbm J}({\boldsymbol\lambda})$ depending on an $N-1$-dimensional
vector ${\boldsymbol\lambda}\in\Lambda$. The $M$-dimensional
ground states ${\mathbf s}$ lives in the eigenspace of ${\mathbbm J}({\boldsymbol\lambda})$ corresponding to the minimal
eigenvalue $j_{min}(\hat{\boldsymbol\lambda})$ and a unique point $\hat{\boldsymbol\lambda}\in\Lambda$ where $j_{min}(\hat{\boldsymbol\lambda})$
assumes its maximum. They can be further characterized by the ``degree of additional degeneracy" $d$ and ``co-degree" $p$
such that $p+d=\frac{1}{2}M(M+1)$. \\
Then, in section \ref{sec:DGS} we present the ``Gram set approach" to the ground state problem that yields further insights. Each spin
configuration ${\mathbf s}$ can also be described by its Gram matrix $G({\mathbf s})$ such that different spin configurations
that are equivalent w.~r.~t.~rotations/reflections yield the same Gram matrix and vice versa. The convex set of all such
$N\times N$-Gram matrices is denoted by ${\mathcal G}_N$. The energy function can be written as an affine functional
${\mathcal G}_N\longrightarrow {\mathbbm R}$ defined by $G\mapsto \mbox{Tr } G\,{\mathbbm J}$. Hence the set of ground states
of ${\mathbbm J}$ corresponds to the face $F$ of ${\mathcal G}_N$ where $G\mapsto\mbox{Tr } G\,{\mathbbm J}$ assumes its minimum.
The afore-mentioned degree $d$ equals the dimension of $F$.
This leads to a more global angle of view: Instead of asking, which are the ground states of a given ${\mathbbm J}$, we rather
analyze the faces $F$ of ${\mathcal G}_N$ and look for the set of matrices ${\mathbbm J}$ such that $G\mapsto\mbox{Tr } G\,{\mathbbm J}$ assumes
its minimum at $F$.

Further topics of this section are the definition of symmetries of the Gram set that are generated by permutations
of spins and partial reflections and the geometry of ``Ising matrices", i.~e., those Gram matrices that correspond to
one-dimensional spin configurations.\\

Next we apply the combined Lagrange variety and Gram set approach to the ground state problem of the general classical
spin triangle, i.~e., for $N=3$. Here it is possible to visualize the Gram set and its faces in a $3$-dimensional plot.
The ground state problem can be explicitly solved for $N=3$. That means that given the three coupling constants one can
determine the corresponding face $F$ and calculate all ground states ${\mathbf s}$ such that $G({\mathbf s})\in F$.
In the Appendix \ref{sec:ES} we calculate all completely elliptic subspaces of ${\mathbbm R}^3$. This set is in $1:1$ correspondence
to the set of closed faces of ${\mathcal G}_3$. A second Appendix \ref{sec:SY} is devoted to the proof that the above-mentioned
symmetries of the Gram set are the only ones. We close with a summary and outlook.


\section{General definitions and results}\label{sec:D}


\subsection{Lagrange variety approach}\label{sec:DLV}


The classical phase space ${\mathcal P}_M$ for the systems of $N$ spins under consideration
consists of all configurations of $M$-dimensional spin vectors
\begin{equation}\label{D1}
\mathbf{s}_\mu,\;\mu=1,\ldots,N
\;,
\end{equation}
subject to the constraints
\begin{equation}\label{D2}
\mathbf{s}_\mu\cdot \mathbf{s}_\mu=1,\;\mu=1,\ldots,N
\;,
\end{equation}
where $\cdot$ denotes the usual inner product of ${\mathbb R}^M$.
Hence  ${\mathcal P}_M$ can be considered as the $N$-fold product of unit spheres,
${\mathcal P}_M={\mathcal S}^{(M-1)}\times\ldots\times {\mathcal S}^{(M-1)}$.
We write the elements ${\mathbf s}$ of ${\mathcal P}_M$  as
$N\times M$-matrices with entries ${\mathbf s}_{\mu,i},\;\mu=1,\ldots,N,\;i=1,\ldots,M$.
Hence ${\mathbf s}$ has $N$ rows ${\mathbf s}_\mu,\,\mu=1,\ldots ,N$ and $M$ columns ${\mathbf s}_i,\,i=1,\ldots,M$.
The ``dimension of ${\mathbf s}$" is defined as its matrix rank, $\mbox{ dim }{{\mathbf s}}\equiv \mbox{ rank }{{\mathbf s}}$.
A subspace $S\subset {\mathbbm R}^N$ is called ``$M$-elliptic" iff there exists an $N\times M$-matrix
${\mathbf s}$ such that its columns  ${\mathbf s}_i\in S$ for $i=1,\ldots,M$ and its rows
${\mathbf s}_\mu,\,\mu=1,\ldots ,N$ satisfy (\ref{D2}). In other words, there exists an $M$-dimensional
spin configuration living on $S$. $S$ is called ``elliptic" if it is $M$-elliptic for some $M>0$
and ``completely elliptic" iff there exists an $N\times M$-matrix
${\mathbf s}$ such that its columns  ${\mathbf s}_i\in S$ for $i=1,\ldots,M$ and its rows
${\mathbf s}_\mu,\,\mu=1,\ldots ,N$ satisfy (\ref{D2}), and moreover, $\mbox{rank } {\mathbf s}=\mbox{ dim } S=M$.
If $S$ is elliptic and $r=\mbox{rank } {\mathbf s}<M$ then we may choose a suitable rotation $R\in O(M)$ such that
${\mathbf s}'={\mathbf s}\,R$ has only $r$ non-zero columns. After deleting the zero columns we obtain
$r=\mbox{rank } {\mathbf s}'=M$. Further restricting $S$ to the subspace $S'$ spanned by the columns of ${\mathbf s}'$
then yields a completely elliptic subspace.

The Heisenberg Hamiltonian $H$ is a smooth function defined on ${\mathcal P}_M$ of the form
\begin{equation}\label{D3}
H(\mathbf{s})=\sum_{\mu,\nu=1}^N J_{\mu \nu}\,\mathbf{s}_\mu\cdot \mathbf{s}_\nu
\;,
\end{equation}
where the coupling coefficients $J_{\mu \nu}$ are the entries of a real, symmetric $N\times N$ matrix $\mathbb J$ with vanishing diagonal.
We will denote by ${\mathcal S}{\mathcal M}(N)$ the linear space of real, symmetric $N\times N$ matrices.
A ``ground state" of the spin system is defined as any configuration $\mathbf s$ where $H({\mathbf s})$ assumes a global minimum
$E_{min}$. In general there exist a lot of ground states. For example, a global rotation/reflection of a ground state is again a ground state
due to the invariance of the Hamiltonian (\ref{D3}) under rotations/reflections. Recall that the group of  all invertible
$M\times M$-matrices $R$ satisfying $R^{-1}=R^\top$ is denoted by $O(M)$.
In general we will call two spin configurations ${\mathbf s}$ and ${\mathbf s}'$ ``$O(M)$-equivalent" iff there exists an $R\in O(M)$
such that ${\mathbf s}'_\mu=R\,{\mathbf s}_\mu$ for all $\mu=1,\ldots,N$.
Further degeneracies
of the ground states apart from the ``trivial" rotational/reflectional degeneracy will be called ``additional degeneracies".

Any ground state ${\mathbf s}$ satisfies the ``stationary state equation" (SSE)
\begin{equation}\label{D4}
  {\mathbb J}({\boldsymbol\kappa})\,{\mathbf s}=0.
\end{equation}
Here ${\mathbb J}({\boldsymbol\kappa})\in {\mathcal S}{\mathcal M}(N){\mathcal S}{\mathcal M}(N)$ is the ``dressed $\mathbb J$-matrix" with entries
\begin{equation}\label{D5}
  \left({\mathbb J}(\boldsymbol\kappa)\right)_{\mu\nu}=J_{\mu \nu}+\kappa_\mu\,\delta_{\mu\nu}
  \;,
\end{equation}
where ${\boldsymbol\kappa}\in{\mathbbm R}^N$ and its components $\kappa_\mu$ are the (negative)
$N$ Lagrange parameters due to the constraints (\ref{D2}).
The ``undressed" matrix ${\mathbb J}$ without ${\boldsymbol\kappa}$ will always denote
a symmetric $N\times N$-matrix with zero diagonal.
As in \cite{S17} we may split the Lagrange parameters  $\kappa_\mu$ into the mean value $\bar{\kappa}$ and the deviations
$\lambda_\mu$ from the mean value according to
\begin{eqnarray}\label{D6a}
  \bar{\kappa} &\equiv& \frac{1}{N} \sum_{\mu=1}^N\,\kappa_\mu,\\
  \label{D6b}
  \lambda_\mu &\equiv& \kappa_\mu-\bar{\kappa}
  \;.
\end{eqnarray}
We denote by $\Lambda$ the $(N-1)$-dimensional subspace of vectors ${\boldsymbol\lambda}\in{\mathbb R}^N$ satisfying
$\sum_{\mu=1}^{N}\lambda_\mu=0$. Then (\ref{D4}) can be written in the form of an eigenvalue equation:
\begin{equation}\label{D7}
{\mathbb J}({\boldsymbol\lambda})\,{\mathbf s}_i=-\bar{\kappa}\,{\mathbf s}_i,\;i=1,\ldots,M.
\end{equation}
The energy (\ref{D3}) does not change if the undressed ${\mathbb J}$-matrix is replaced by ${\mathbb J}({\boldsymbol\lambda})$.
We introduce some more notation. Let $j_\alpha({\boldsymbol\lambda})$ denote the $\alpha-$th
eigenvalue of ${\mathbb J}({\boldsymbol\lambda})$ and
$j_{min}({\boldsymbol\lambda})$ its lowest eigenvalue for ${\boldsymbol\lambda} \in \Lambda$.
Then the following holds:
\begin{theorem}\label{Theorem1}
 Under the preceding definitions there exists a unique point $\hat{\boldsymbol\lambda}\in \Lambda$
 such that $j_{min}({\boldsymbol\lambda})$ assumes  its maximum $\hat{\jmath}$
 at $\hat{\boldsymbol\lambda}$ and the corresponding eigenspace $S$ of ${\mathbb J}(\hat{\boldsymbol\lambda})$
 is elliptic. Moreover, all ground states of (\ref{D3}) live on $S$.
\end{theorem}

This follows from the theorems 1,2,3 of \cite{S17}.

Now consider an $M''$-elliptic subspace $S'$,
for example the eigenspace of $({\mathbb J}(\hat{\boldsymbol\lambda}),\hat{\jmath})$ according to Theorem \ref{Theorem1}.
Let $W'$ denote
an $N\times M'$-matrix such that the $M'$ columns of $W'$ form a basis of $S'$, hence $M'=\mbox{ dim }S'$.
$W'$ can be viewed as an injective linear map $W':{\mathbbm R}^{M'}\longrightarrow {\mathbbm R}^{N}$ and, similarly,
$W'^\top$ as a surjective linear map $W'^\top:{\mathbbm R}^{N}\longrightarrow {\mathbbm R}^{M'}$.

According to the definition of ``$M''$-elliptic" there exists a spin configuration ${\mathbf s}$ such that the
columns ${\mathbf s}_i\in S'$ for $i=1,\ldots, M''$. These columns are unique linear combinations of the basis vectors
$W'_j,\;j=1,\ldots M'$, i.~e.~,
\begin{equation}\label{D8}
 {\mathbf s}_{\mu i} =\sum_{j=1}^{M'}W'_{\mu j}\,\Gamma_{j\,i},\quad i=1,\ldots, M'',\;\mu=1,\ldots,N.
\end{equation}
In matrix notation (\ref{D8}) is written as ${\mathbf s}=W'\;{\boldsymbol\Gamma}$, where ${\boldsymbol\Gamma}$ is the
$M'\times M''$-matrix with entries $\Gamma_{j\,i}$.
The condition (\ref{D2}) that the rows ${\mathbf s}_\mu$ of ${\mathbf s}$ are unit vectors assumes the form
\begin{equation}\label{D9}
1=\left({\mathbf s}\,{\mathbf s}^\top\right)_{\mu\mu} =
 \left(W'\,{\boldsymbol\Gamma}\,{\boldsymbol\Gamma}^\top\,W'^\top\right)_{\mu\mu}
 =\left(W'\,{\boldsymbol\Delta}'\,W'^\top\right)_{\mu\mu} ,
\end{equation}
where  ${\boldsymbol\Delta}'\equiv {\boldsymbol\Gamma}\,{\boldsymbol\Gamma}^\top$ is a positively semi-definite
$M'\times M'$-matrix. Let ${\mathcal S}'$ be the set of solutions ${\boldsymbol\Delta}'\ge 0$ of (\ref{D9}).
According to the assumption of ellipticity this solution set is not empty. Generally, ${\mathcal S}'$ is the
intersection of the affine space of solutions of (\ref{D9}) with the closed convex cone ${\mathcal S}{\mathcal M}_+(M')$
of positively semi-definite $M'\times M'$-matrices, and hence a closed convex set. Let $F={\sf face}\left({\mathcal S}'\right) $, the face of
${\mathcal S}{\mathcal M}_+(M')$ generated by the subset ${\mathcal S}'$. According to Theorem \ref{Theorem1}, see section
\ref{sec:DGS}, $F$ is of the form ${\sf fac}(U)$, where $U$ is a linear subspace of ${\mathbbm R}^{M'}$.
and ${\sf fac}(U)$ is the face of all
positively semi-definite $M'\times M'$-matrices that live on $U$.
Let  $M=\mbox{ dim }U$. Further define $S\equiv W'[U]\equiv\{W'\,u\left| u\in U\right.\}$.
Here $W'$ is considered as an injective linear map, see above, hence $\mbox{dim }S=\mbox{dim }U=M$.
Without loss of generality we may assume that the first
$M$ columns of $W'$ form a basis of $S$. Let $W$ denote the corresponding $N\times M$-submatrix of $W'$,
and ${\boldsymbol\Delta}$ the correspondingly truncated $M\times M$-submatrix of ${\boldsymbol\Delta}'$.
Then (\ref{D9}) assumes the form
\begin{equation}\label{D10}
1=\left(W\,{\boldsymbol\Delta}\,W^\top\right)_{\mu\mu}.
\end{equation}
Let ${\mathcal S}$ denote the set of solutions ${\boldsymbol\Delta}\ge 0$ of (\ref{D10}).
 It is essentially the same
set as ${\mathcal S}'$ but now the solutions are considered as $M\times M$-matrices.
Usually ${\mathcal S}$ is represented by the convex range ${\mathcal S}_{par}$ of suitable parameters.
If all solutions ${\boldsymbol\Delta}\in {\mathcal S}$  would have  $\mbox{ rank } {\boldsymbol\Delta}<M$ then
they would generate a face of ${\mathcal S}{\mathcal M}_+(M)$ that is smaller than $F={\sf fac}(U)$ which contradicts the
definition of $F$. Hence there exists some ${\boldsymbol\Delta}_0\in {\mathcal S}$ with $\mbox{ rank } {\boldsymbol\Delta}_0=M$
and consequently ${\boldsymbol\Delta}_0>0$. It follows that ${\mathbf s}\equiv W\,\sqrt{{\boldsymbol\Delta}_0}$ is
a spin configuration, i.~e., satisfying (\ref{D2}), with $\mbox{ rank } {\mathbf s}=\mbox{ dim }{\mathbf s}=M$.
(For the proof of the latter statement, let $x\neq 0$ be a vector satisfying ${\mathbf s}\,x=0$, then
$y\equiv \sqrt{{\boldsymbol\Delta}_0}\,x\neq 0$ satisfies $W\,y=0$, i.~e.~, the columns of $W$ are linearly dependent. This is
a contradiction to the assumption that the columns of $W$ form a basis of $S$.)

In this way we have shown that every elliptic subspace $S'\subset {\mathbbm R}^N$ contains a completely elliptic
subspace $S\subset S'$. In section \ref{sec:ST} we will give an example of $S$ being a proper subspace of $S'$
in the context of the spin triangle.
In accordance with \cite{S17} we will call (\ref{D10}) together with the condition ${\boldsymbol\Delta}\ge 0$
the ``additional degeneracy equation" (ADE). It has been shown in \cite{S17} that its solution set ${\mathcal S}={\mathcal S}_{ADE}$ is of the form
${\mathcal S}=\left(\Delta_0+P^\perp\right) \cap {\mathcal S}{\mathcal M}_+(M)$. Here
$P$ is the subspace of ${\mathcal S}{\mathcal M}(M)$ spanned by the the rank $1$ matrices
$P_\mu,\mu=1,\ldots,N,$. They are defined as the projectors onto the $1$-dimensional subspaces
spanned by the $\mu$-th row $W_\mu$ of $W$ multiplied by $\| W_\mu\|^2$.
The orthogonal complement $P^\perp$ of $P$ is defined w.~r.~t.~the inner product
$\langle A|B\rangle \equiv \mbox{Tr}\left( A\,B\right)$ of ${\mathcal S}{\mathcal M}(M)$.
It follows \cite{S17}
that ${\mathcal S}$ is a $d$-dimensional closed convex set where the ``degree $d$"
of $S$ is defined by $d\equiv  M(M+1)/2\,-\,p$, and the ``co-degree $p$" of $S$
by $p\equiv\mbox{dim } P$.

We stress that the solution set ${\mathcal S}$ depends on the choice of the matrix $W$ and the latter is, in general, not unique.
The condition that the $M$ columns of $W$ form a basis of $S$ allows that $W$ can be replaced by $W\,A$, where
$A$ is some invertible  $M\times M$-matrix. This induces the transformation ${\boldsymbol \Delta}\mapsto A\,{\boldsymbol \Delta}\,A^{-1}$
of ${\mathcal S}$ which is an affine bijection ${\mathcal S}\longrightarrow \widetilde{\mathcal S}$.


\subsection{Gram set approach}\label{sec:DGS}

\subsubsection{Generalities}\label{sec:DGSG}

We will recapitulate and generalize some notions already introduced in \cite{SL03} and \cite{S17}.
For each spin configuration ${\mathbf s}$ we define the ``Gram matrix" $G=G({\mathbf s})\equiv {\mathbf s}\,{\mathbf s}^\top$
with entries $G_{\mu\nu}={\mathbf s}_\mu\cdot{\mathbf s}_\nu$.
Hence $G$ will be a symmetric $N\times N$-matrix that is positively semi-definite, $G\ge 0$, and satisfies $G_{\mu\mu}=1$
for all $\mu=1,\ldots,N$. Moreover, $\mbox{rank }(G)=\mbox{rank }({\mathbf s})\le N$.

Conversely, if $G$ is a positively semi-definite $N\times N$-matrix with $\mbox{rank }G=M\le N$, satisfying $G_{\mu\mu}=1$
for all $\mu=1,\ldots,N$, then ${\mathbf s}\equiv \sqrt{G}$ defines a spin configuration such that $G=G({\mathbf s})$.
The correspondence between spin configurations ${\mathbf s}$ and Gram matrices is many-to-one:
Let $R\in O(M)$ be a rotation/reflection,
then the two configurations ${\mathbf s}_\mu$ and $R\,{\mathbf s}_\mu,\;\mu=1,\ldots,N,$ will obviously yield the same Gram matrix.
Conversely, if $G({\mathbf s})=G({\mathbf s}')$ then ${\mathbf s}$ and ${\mathbf s}'$ will be $O(M)$-equivalent, see Proposition $4$ in \cite{S17}.
Hence the representation of spin configurations by Gram matrices exactly
removes the ``trivial" rotational/reflectional degeneracy of possible ground states. If there is no danger of confusion
we will also refer to the Gram matrices as ``states".
If ${\mathbf s}$ runs through all $N$-dimensional unit vector spin configurations the corresponding set of Gram matrices will be denoted
by ${\mathcal G}_N$ or simply by ${\mathcal G}$ if the dimension $N$ is obvious from the context.
More formally, we define the ``Gram set" ${\mathcal G}_N$ as the set of all positively semi-definite $N\times N$-matrices $G$,
satisfying $G_{\mu\mu}=1$ for all $\mu=1,\ldots,N$. It satisfies:
\begin{lemma}\label{Lemma6}
${\mathcal G}$ is a compact convex subset of ${\mathcal S}{\mathcal M}(N)$.
\end{lemma}
The convexity follows immediately since ${\mathcal G}$ can be viewed as the intersection
of the convex cone of all positively semi-definite $N\times N$-matrices with the $N$ affine hyperplanes defined by
$G_{\mu\mu}=1$. It can also easily be shown that ${\mathcal G}$ is closed and bounded, hence compact.
The interior of ${\mathcal G}$ is formed by the open subset of Gram matrices with rank $N$ and its boundary
by the Gram matrices with rank $M<N$. Especially, if $N>3$ the ``physical" spin configurations satisfying $\mbox{rank}(G)\le 3$
will form a non-convex subset of the boundary of ${\mathcal G}$.\\

We now write the Hamiltonian (\ref{D3}) as
\begin{equation}\label{DGS1}
H({\mathbf s})=\sum_{\mu,\nu=1}^{N} J_{\mu\nu}{\mathbf s}_\mu\cdot{\mathbf s}_\nu=\sum_{\mu,\nu=1}^{N} J_{\mu\nu}G_{\mu\nu}={\mbox Tr}({\mathbb J}\,G)
\;.
\end{equation}
Hence the energy can be viewed as an affine functional \\
${\mathcal J}:{\mathcal G}\longrightarrow{\mathbb R},\,G\mapsto {\mbox Tr}({\mathbb J}\,G)$ defined on the Gram set.
Similarly as in section \ref{sec:DLV} all matrices ${\mathbb J}({\boldsymbol \lambda}),\;{\boldsymbol \lambda}\in\Lambda$ that differ only in the
parameters ${\boldsymbol\lambda}$  generate the same functional ${\mathcal J}$. According to general theorems, an affine functional ${\mathcal J}$
defined on a compact convex set ${\mathcal G}$ assumes its minimum at some closed face ${\mathcal F}$ of ${\mathcal G}$. Recall that a ``face" of
${\mathcal G}$ is a convex subset ${\mathcal F}\subset{\mathcal G}$
such that for all $x\in{\mathcal F},\; y,z \in{\mathcal G}$ and $0<\alpha <1$ the equation
$x=\alpha\,y+(1-\alpha)\,z$ implies $y,z\in{\mathcal F}$.
If ${\mathcal F}$ consists of a single point, ${\mathcal F}=\{ f\}$, then $f$ is called an ``extremal point" of  ${\mathcal G}$.

These facts motivate a closer investigation of $\mbox{\sf Fac}({\mathcal G})$, the set of closed faces of ${\mathcal G}$. Let ${\mathcal K}$ be the
larger convex set of real, symmetric $N\times N$ matrices $W\ge 0$ satisfying $\mbox{Tr}\, W=N$. If we replace ``real, symmetric" by "Hermitean"
and ``$\mbox{Tr}\, W=N$" by ``$\mbox{Tr}\, W=1$" we obtain the convex set of (mixed and pure) states of a finite-dimensional quantum system.
A closed face of this set consists of all density matrices that ``live" on a linear subspace  $S\subset {\mathbb C}^N$ of the finite-dimensional
Hilbert space ${\mathbb C}^N$, i.~e., the eigenvectors of $W$ corresponding to positive eigenvalues lie in $S$, see, e.~g., \cite{BZ06} Theorem 8.3,
or \cite{L83} Theorem 6.2 for the infinite-dimensional case.
This result can be transferred to the above-defined convex set ${\mathcal K}$ and yields the criterion that a closed face of ${\mathcal K}$ consists
of all matrices  $W\in{\mathcal K}$ that live on a certain linear subspace $S\subset{\mathbb R}^N$.
We will denote this face by $\mbox{\sf fac}(S)$.
Let ${\mathbb G}$ be the intersection of the
$N$ affine hyperplanes in the linear space ${\mathcal S}{\mathcal M}(N)$ defined by $G_{\mu\mu}=1,\;\mu=1,\ldots,N$ such that
\begin{equation}\label{DGS2}
 {\mathcal G}={\mathbb G}\cap{\mathcal K}
 \;.
\end{equation}
Then we have the following situation: Every face ${\mathcal F}$ of ${\mathcal G}$ can be extended to a face $\overline{\mathcal F}$
of ${\mathcal K}$, namely the intersection of all faces of ${\mathcal K}$ containing ${\mathcal F}$. Conversely,
every face ${\sf F}$ of ${\mathcal K}$ can be restricted to the face $\underline{{\sf F}}\equiv {\sf F}\cap{\mathbb G}$.
It can be shown that the restriction of the extension returns the original face:
\begin{lemma}\label{Lemma7}
With the preceding definitions, $\underline{\overline{\mathcal F}}=\overline{\mathcal F}\cap{\mathbb G}=
{\mathcal F}$ for all ${\mathcal F}\in\mbox{\sf Fac}({\mathcal G})$.
\end{lemma}
From this lemma it follows that the map ${\mathcal F}\mapsto \overline{\mathcal F}$ is injective and that $\mbox{\sf Fac}({\mathcal G})$ is in $1:1$
correspondence with a certain subset ${\mathbb E}\subset\mbox{\sf Fac}({\mathcal K})$, which is the image of the above extension map.\\

\noindent{\bf Proof of Lemma \ref{Lemma7}}\\
Obviously, ${\mathcal F}\subset \overline{\mathcal F}\cap{\mathbb G}\subset{\mathcal G}$ .
It is easily shown that the extension of the face ${\mathcal F}$ can be obtained as
\begin{eqnarray}\nonumber
\overline{\mathcal F}&=&\left\{ f_1\in{\mathcal K}\,|\, f=\alpha f_1 +(1-\alpha)f_2,\right. \\
\label{AP20a}
&&\left. \mbox{ where }f\in{\mathcal F},\,f_2\in{\mathcal K},\,0<\alpha<1\right\}
\;.
\end{eqnarray}
In order to show $\overline{\mathcal F}\cap{\mathbb G}\subset{\mathcal F}$
let $f_1\in \overline{\mathcal F}\cap{\mathbb G}\subset {\mathcal K}\cap{\mathbb G}={\mathcal G}$
and consider the equation  $f=\alpha f_1 +(1-\alpha)f_2$ in (\ref{AP20a}).
We conclude $f\in{\mathcal F}\subset{\mathcal G}\subset{\mathbb G}$ and hence also $f_2\in {\mathbb G}$ and $f_2\in {\mathcal G}$.
Thus we have $f=\alpha f_1 +(1-\alpha)f_2$ where $f\in{\mathcal F},\, f_1,f_2\in{\mathcal G}$ and $0<\alpha<1$. This implies
$f_1\in{\mathcal F}$ since ${\mathcal F}$ is a face of ${\mathcal G}$ and the proof of lemma \ref{Lemma7} is completed.
 \hfill$\Box$\\

There are obvious connections between the Lagrange variety approach and the Gram set approach that will be summarized in the following
theorems:
\begin{theorem}\label{Theorem1}
Let ${\mathcal F}\equiv\{ G\in{\mathcal G}\left| \mbox{Tr}\left({\mathbbm J}\,G\right)=E_{min}\right.\}$. Then ${\mathcal F}$ is a closed face
of $\mathcal G$. Let $\overline{{\mathcal F}}\in{\sf Fac}\left({\mathcal K}\right)$ denote its extension. It will be of the form
$\overline{{\mathcal F}}= {\sf fac}(S)$ for some $M$-dimensional subspace $S$ of ${\mathbbm R}^N$. Let $W$ be an $N\times M$-dimensional
matrix such that its $M$ columns form a basis of $S$ and ${\mathcal S}$ denote the solution set of the ADE (\ref{D10})
together with the condition ${\boldsymbol\Delta}\ge 0$. Then the map
\begin{eqnarray}\label{DGS3a}
  w&:& {\mathcal S}\longrightarrow {\mathcal F}\\
  && {\boldsymbol\Delta} \mapsto W\, {\boldsymbol\Delta} \,W^\top
\end{eqnarray}
will be an affine bijection, and hence the two convex sets ${\mathcal S}$ and ${\mathcal F}$ are affinely isomorphic.
Moreover, $S$ is completely elliptic.

 The above-defined subset ${\mathbb E}\subset\mbox{\sf Fac}({\mathcal K})$ consists
of all faces ${\sf F}$ of ${\mathcal K}$ of the form ${\sf F}={\sf fac}\left(S\right)$ where $S$ is completely elliptic.
\end{theorem}

\begin{theorem}\label{Theorem1a}
(i) If $\check{\boldsymbol\kappa}$ denotes the Lagrange parameters of some ground state ${\mathbf s}$ of ${\mathbbm J}$
with Gram matrix $G=G({\mathbf s})$ then ${\mathbbm J}(\check{\boldsymbol\kappa})$ and $G$ have orthogonal supports, i.~e.,
${\mathbbm J}(\check{\boldsymbol\kappa})\,G=0$.

(ii) Conversely, let a non-trivial closed face ${\mathcal F }$ of ${\mathcal G}$ of the form ${\mathcal F }={\sf fac}(S)$ be given,
where $S$ is a completely elliptic subspace of ${\mathbbm R}^N$ and $S^\perp$ denotes its orthogonal complement.
Then each positively semi-definite matrix $A\ge 0$ with support $S^\perp$ can be written as $A={\mathbbm J}({\boldsymbol\kappa})$
such that every $G=G(\mathbf s)\in{\mathcal F}$ yields a ground state of ${\mathbbm J}({\mathbf 0})$ with Lagrange parameters ${\boldsymbol\kappa}$.

(iii) For every $G_0$ at the boundary of ${\mathcal G }$ let ${\mathcal F}_0$ be the closed face generated by $G_0$.
It is of the form ${\mathcal F}_0={\sf fac}(S)$, where $S$ is a completely elliptic subspace of ${\mathbbm R}^N$.
We denote by ${\mathcal C}(G_0)$ the convex cone consisting of all linear functionals that are non-negative on ${\mathcal G}-G_0$. Then the map
$A\mapsto f(A)$ with $f(A)(G-G_0)=\mbox{ Tr }\left(A(G-G_0)\right)$ defines an affine bijection between the
cone of positively semi-definite matrices $A\ge 0$ with support $S^\perp$ and ${\mathcal C}(G_0)$.
\end{theorem}

\noindent{\bf Proof of theorem \ref{Theorem1a}}\\
The first claim (i) follows from ${\mathbbm J}(\check{\boldsymbol\kappa})\,{\mathbf s}=0$, cf.~(\ref{D4}),
which implies ${\mathbbm J}(\check{\boldsymbol\kappa})\,G={\mathbbm J}(\check{\boldsymbol\kappa})\,{\mathbf s}\,{\mathbf s}^\perp=0$.
For the second part (ii) we note that, according to the assumptions, ${\mathbbm J}({\boldsymbol\kappa})\,G=0$ holds
which implies
\begin{eqnarray}
\label{proof1a}
 B\,B^\top&\equiv&\left({\mathbbm J}({\boldsymbol\kappa})\,{\mathbf s}\right)\,\left({\mathbbm J}({\boldsymbol\kappa})\,{\mathbf s}\right)^\top \\
 \label{proof1b}
   &=& {\mathbbm J}({\boldsymbol\kappa})\,{\mathbf s}\,{\mathbf s}^\top\,{\mathbbm J}({\boldsymbol\kappa})^\top\\
   \label{proof1c}
   &=& {\mathbbm J}({\boldsymbol\kappa})\,G\,{\mathbbm J}({\boldsymbol\kappa})^\top=0
   \;,
\end{eqnarray}
and hence $B={\mathbbm J}({\boldsymbol\kappa})\,{\mathbf s}=0$. This is the SSE (\ref{D4}). ${\mathbf s}$ is a ground state
of ${\mathbbm J}({\mathbf 0})$ since ${\mathbbm J}({\boldsymbol\kappa})\ge 0$ implies that the negative mean value $-\overline{\kappa}$
is the lowest eigenvalue of ${\mathbbm J}({\boldsymbol\lambda})$.

From this (iii) follows immediately.   \hfill$\Box$\\

\vspace{4cm}
\begin{figure}[ht]
  \centering
    \includegraphics[width=0.3\linewidth,viewport =40mm 0mm 80mm 80mm]{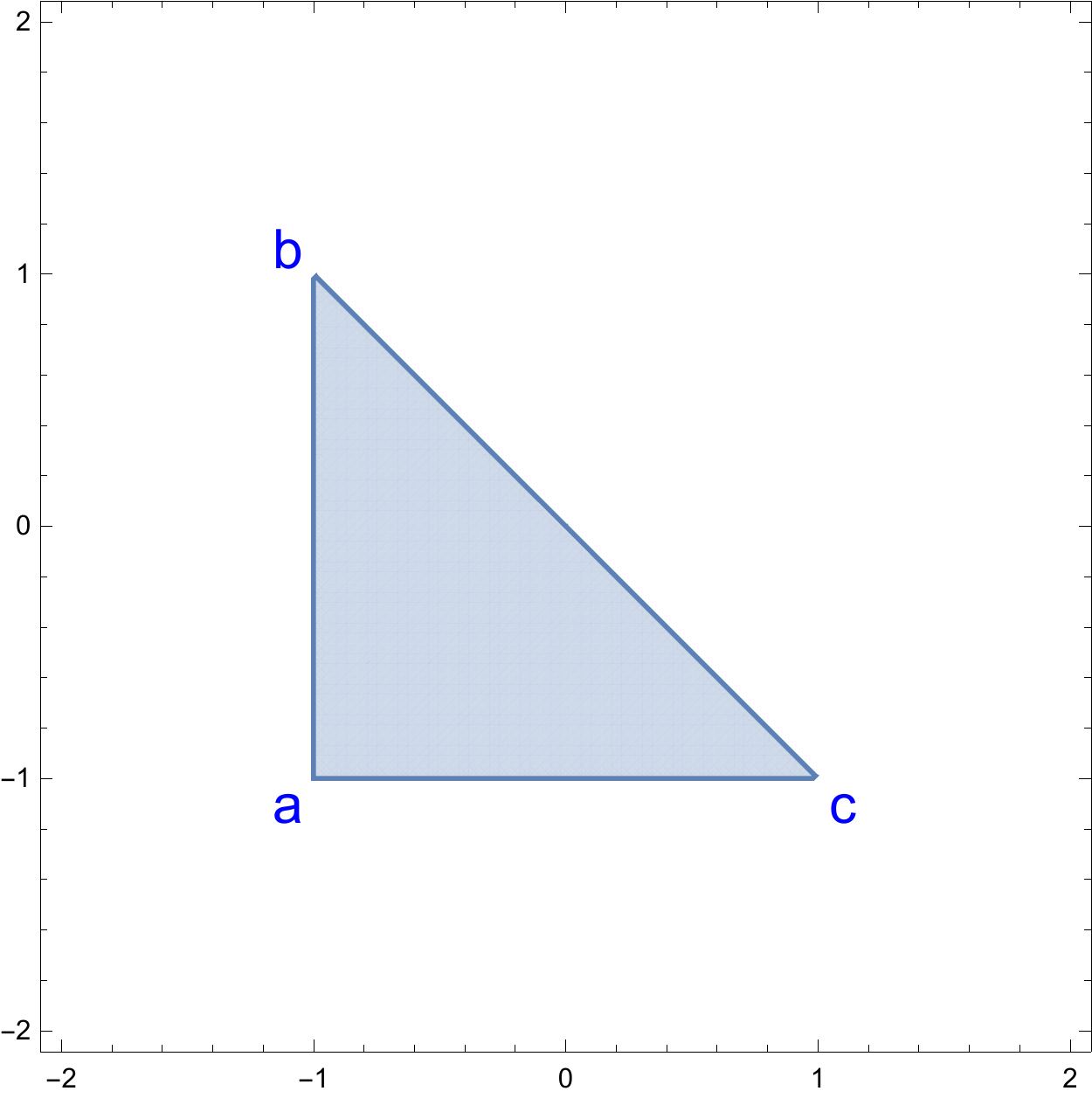}
  \caption[Region plot]
  {The convex region ${\mathcal S}_{par}$ of parameters $(x,y)$ corresponding to the ground states of the uniform
  AF spin tetrahedron. The extremal points $\mathbf{a,b}$ and ${\mathbf c}$ correspond to the Ising states $\uparrow \downarrow\downarrow\uparrow$,
  $\uparrow \downarrow\uparrow\downarrow$, and $\uparrow \uparrow\downarrow\downarrow$, resp.~.
  }
  \label{FIGRP}
\end{figure}

We will close this section with a simple example that illustrates some of the points discussed above although it is almost trivial considered  as a pure ground state problem. It is the spin tetrahedron with uniform AF coupling according to the Hamiltonian
\begin{equation}\label{ex1}
H={\mathbf s}_1\cdot {\mathbf s}_2+{\mathbf s}_1\cdot {\mathbf s}_3+{\mathbf s}_1\cdot {\mathbf s}_4
+{\mathbf s}_2\cdot {\mathbf s}_3+{\mathbf s}_2\cdot {\mathbf s}_4+{\mathbf s}_3\cdot {\mathbf s}_4
\;.
\end{equation}

Since $H=\frac{1}{2}  \left({\mathbf S}^2-4\right)$ where ${\mathbf S}=\sum_{\mu=1}^{4}{\mathbf s}_\mu$ is the total spin,
it follows that the energy has the minimal value $E_{min}=-2$ for any spin configuration with ${\mathbf S}={\mathbf 0}$.
Such spin configurations exist with any dimension between $1$ and $3$. The eigenvalues of the undressed ${\mathbbm J}$-matrix
are $j_{max}=\frac{3}{2}$ and $j_{min}=-\frac{1}{2}$, the latter being $3$-fold degenerate. The above-mentioned ground states
live on the $3$-dimensional eigenspace $S$ corresponding to $j_{min}=-\frac{1}{2}$, which is hence completely elliptic.
Their Gram matrices $G(x,y)$ form a two-dimensional
face ${\mathcal F}$ of ${\mathcal G}_4$ parametrized by $(x,y)\in{\mathcal S}_{par}$, see Figure \ref{FIGRP}. They are given by
$G(x,y)=$
\begin{equation} \label{ex2}
 \left(
\begin{array}{cccc}
 1 & x & y & -x-y-1 \\
 x & 1 & -x-y-1 & y \\
 y & -x-y-1 & 1 & x \\
 -x-y-1 & y & x & 1 \\
\end{array}
\right)\;.
\end{equation}
The extremal points $\mathbf {a,b,c}$ of ${\mathcal S}_{par}$ correspond to the three Ising states $\uparrow \downarrow\downarrow\uparrow$,
$\uparrow \downarrow\uparrow\downarrow$, and $\uparrow \uparrow\downarrow\downarrow$ with total spin ${\mathbf S}={\mathbf 0}$.
The three one-dimensional faces joining two points of $\mathbf {a,b,c}$, resp., correspond to co-planar ground states generated
by the independent rotation of two pairs of Ising states of the form $\uparrow\downarrow$. Finally, the interior of the
face ${\mathcal F}$ corresponds to $3$-dimensional spin configurations with ${\mathbf S}={\mathbf 0}$ which show an additional
degeneracy that can be described by two parameters, e.~g., the scalar products $x={\mathbf s}_1\cdot {\mathbf s}_2={\mathbf s}_3\cdot {\mathbf s}_4$
and $y={\mathbf s}_1\cdot {\mathbf s}_3={\mathbf s}_2\cdot {\mathbf s}_4$, see (\ref{ex2}).

\vspace{3cm}
\begin{figure}[ht]
  \centering
    \includegraphics[width=0.3\linewidth,viewport =40mm 0mm 80mm 80mm]{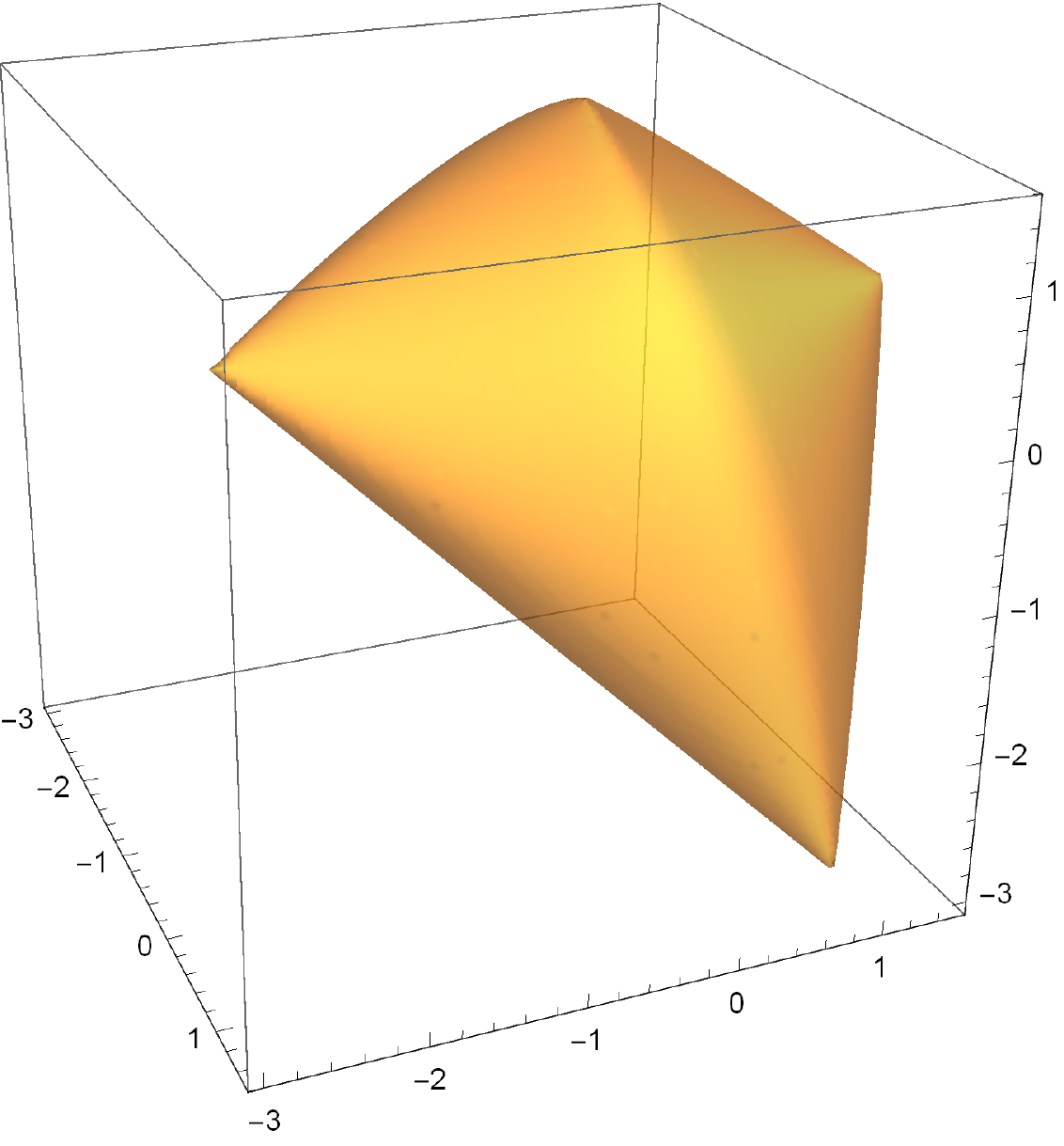}
  \caption[Region plot]
  {The convex set representing the intersection of the $4$-dimensional cone ${\mathcal C}$ with the plane $\xi=1$.
  The cone ${\mathcal C}$ approximates the Lagrange variety $j_{min}({\boldsymbol\lambda})$ in the neighborhood
  of the Lagrange parameters ${\boldsymbol\lambda}={\mathbf 0}$ corresponding to the ground states of the uniform
  AF spin tetrahedron.
  }
  \label{FIGC1}
\end{figure}

It is interesting to visualize the neighborhood of the Lagrange variety $0=\det \left({\mathbbm J}({\boldsymbol\lambda})+\overline{\kappa}{\mathbbm 1}\right)$
at the point ${\boldsymbol\lambda}={\mathbf 0},\;\overline{\kappa}=\frac{1}{2}$. As expected from the general theory
of \cite{S17} it is given by a vertical convex cone ${\mathcal C}$, albeit not a circular one.
The reason for this is that the Taylor expansion of the determinant only starts with terms of third order
in the variables $\lambda_1,\,\lambda_2,\,\lambda_3$ and $\xi\equiv\kappa-\overline{\kappa}$. Since it is not possible to
visualize a cone in four dimensions we rather display its intersection with the plane $\xi=1$,see Figure \ref{FIGC1}.
This is a convex set with the analytical expression
\begin{equation}\label{ex3a}
  4 -p_1-2\, s_1^2+2\,s_2+s_1\,s_2 \ge  0\;,
\end{equation}
where the symmetrical polynomials
\begin{eqnarray}
 \label{ex3b}
  s_1 &\equiv& \lambda_1+\lambda_2+\lambda_3 \\
  \label{ex3c}
  s_2 &\equiv& \lambda_1\lambda_2+\lambda_2\lambda_3+\lambda_1\lambda_3\\
  \label{ex3d}
  p_1&\equiv & \lambda_1\,\lambda_2\,\lambda_3
 \end{eqnarray}
have been used.

\subsubsection{Symmetries}\label{sec:DGSS}

The Gram set ${\mathcal G}$ has a large finite symmetry group ${\sf Sym}$. To determine it we first consider permutations $\pi$ of
$N$ spins, $\pi\in S_N$. They act in a natural way on spin configurations by ${\mathbf s}_\mu \mapsto {\mathbf s}_{\pi(\mu)},\;\mu=1,\ldots,N$
and on Gram matrices by $G\mapsto G',\,G'_{\mu,\nu}={\mathbf s}_{\pi(\mu)}\cdot {\mathbf s}_{\pi(\nu)} = G_{\pi(\mu),\pi(\nu)}$.
If $U(\pi)$ is the linear $O(N)$-representation of $\pi\in S_N$, given by $U(\pi)_{\mu,\nu}=\delta_{\mu,\pi(\nu)},\;\mu,\nu=1,\ldots,N$
then the above action can be written as
\begin{equation}\label{DSG7}
 G\mapsto G'=\Pi(G)\equiv U(\pi)^{-1} \,G\,U(\pi)
 \;.
\end{equation}
It follows that $G\ge 0$ iff $G'\ge 0$ and that $G_{\mu\mu}=1$ for all $\mu=1,\ldots,N$ iff $G'_{\mu\mu}=1$ for all $\mu=1,\ldots,N$.
This means that $\Pi$, defined in (\ref{DSG7}), is an affine bijection of ${\mathcal G}$ onto itself, hence a symmetry.
We will denote the group of symmetries $\Pi$ generated by permutations $\pi\in S_N$ by ${\sf Per}$.\\
Secondly, we consider the partial reflections $\rho_{\mathcal T}$ defined by
\begin{equation}\label{DSG8}
{\mathbf s}_\mu\mapsto -{\mathbf s}_\mu \mbox{ for }\mu\in{\mathcal T},\;
{\mathbf s}_\mu\mapsto {\mathbf s}_\mu \mbox{ for }\mu\notin{\mathcal T}
\;,
\end{equation}
where ${\mathcal T}$ is an arbitrary subset of $\{1,\ldots,N\}$. There is an obvious representation
of  $\rho_{\mathcal T}$ by diagonal matrices $V(\rho_{\mathcal T})\in O(N)$ with entries
$V(\rho_{\mathcal T})_{\mu\mu}=-1$ iff  $\mu\in{\mathcal T}$
and $+1$ iff $\mu\notin{\mathcal T}$. Obviously, $V(\rho_{\mathcal T})^2={\mathbbm 1}_N$.
We define ${\sf R}_{\mathcal T}(G)\equiv V(\rho_{\mathcal T})\,G\,V(\rho_{\mathcal T})$ and denote
the Abelian group of transformations ${\sf R}_{\mathcal T}$ of ${\mathcal G}$ by ${\sf Rf}$.
This group is generated by the subset
of partial reflections of just one spin, i.~e., ${\mathcal T}=\{\alpha\},\;1\le\alpha\le N$. The latter act on Gram matrices
by inverting the signs of the $\alpha$th row and the $\alpha$th column, leaving $G_{\alpha\alpha}=1$ unchanged:
\begin{eqnarray}\label{DSG9a}
G\mapsto \tilde{G}&=&V(\rho_{\{\alpha\}})\,G\,V(\rho_{\{\alpha\}}) ,\\
\label{DSG9b}
\tilde{G}_{\mu\nu}&=&(-1)^{\delta_{\mu\alpha}}\,(-1)^{\delta_{\nu\alpha}}\,G_{\mu\nu}
\;.
\end{eqnarray}
Obviously the transformation $G\mapsto \tilde{G}$ leaves $\det(G)$ invariant, and, more general, also the principal minors of $G$.
Hence, by Sylvester's criterion, $G\ge 0$ iff $\tilde{G}\ge 0$ and also the partial reflections generate symmetries ${\sf R}_{\mathcal T}$ of ${\mathcal G}$.
\\
Let ${\sf Sym}$ be the group of all bijections of ${\mathcal G}$ generated by ${\sf Per}$ and ${\sf Rf}$.
It seems plausible
and will be proven in the Appendix \ref{sec:SY} that
${\sf Sym}$ is the group of all affine bijections of ${\mathcal G}$. At first sight one might think
that there are $N!\,2^N$ transformations in ${\sf Sym}$. However the order of ${\sf Sym}$ is only $N!\,2^{N-1}$ since
$({\mathbf s}_\mu)_{\mu=1,\ldots,N}$ and $(-{\mathbf s}_\mu)_{\mu=1,\ldots,N}$ generate the same Gram matrix.\\
The action of the group ${\sf Sym}$ can also be extended to the set of symmetric $N\times N$ matrices ${\mathbb J}$ with vanishing diagonal:
$\gamma\in{\sf Sym}$ acts on ${\mathbb J}$ in the same way as on $G$ leaving the diagonal unchanged. It follows that
\begin{equation}\label{DSG10}
 \mbox{Tr } (G\, {\mathbb J}) =  \mbox{Tr } (\gamma(G)\, \gamma({\mathbb J}))
 \;.
\end{equation}
Hence if $G$ represents a ground state of ${\mathbb J}$ then $\gamma(G)$ represents a ground state of $\gamma({\mathbb J})$ with
the same energy $E_{min}$. This considerably simplifies the investigation of ground states, as we will see in section \ref{sec:ST}.\\

\subsubsection{Ising matrices}\label{sec:DGSI}

\vspace{2cm}
\begin{figure}[ht]
  \centering
    \includegraphics[width=0.3\linewidth,viewport =40mm 0mm 80mm 80mm]{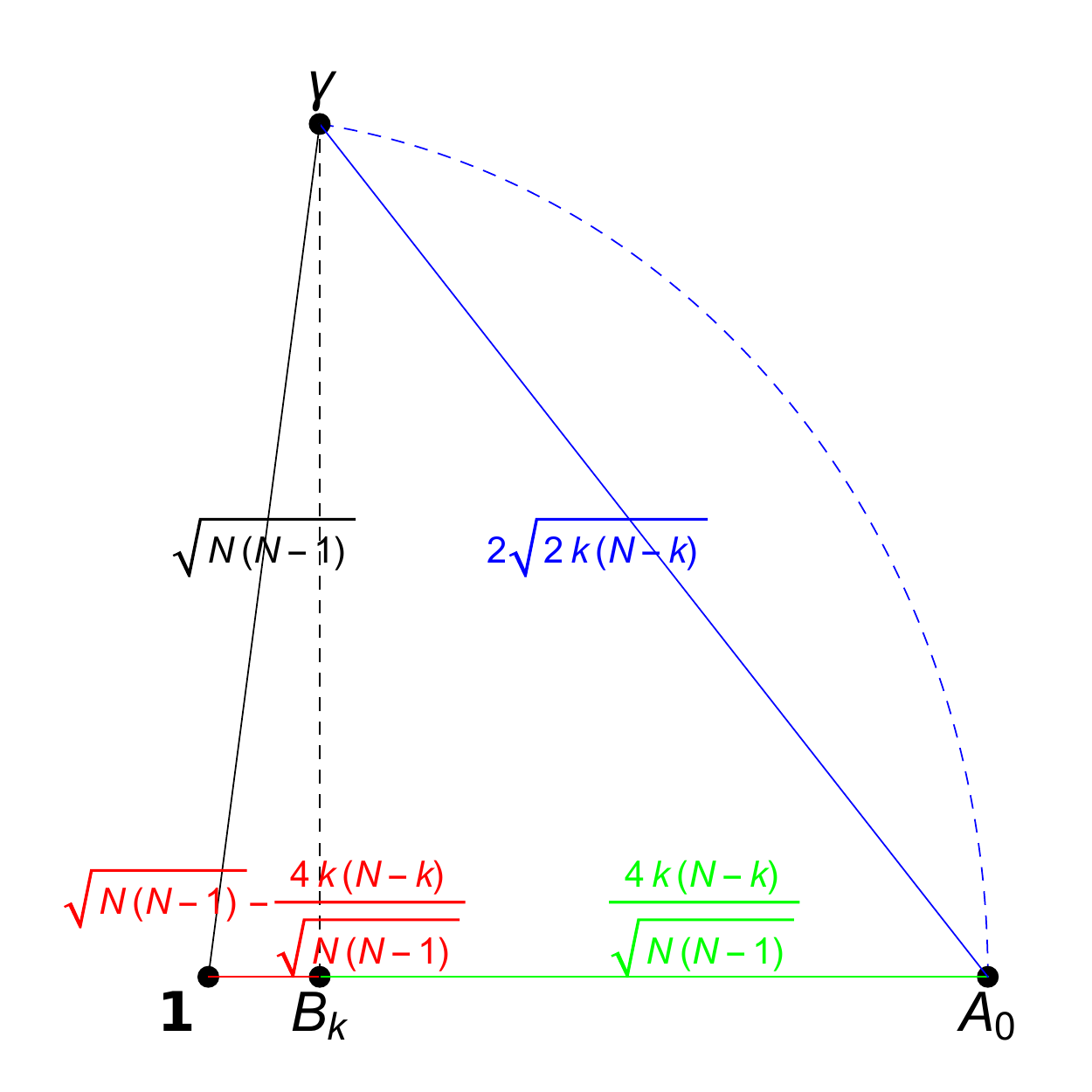}
  \caption[Ising]
   {Illustration of the geometry of Ising matrices according to Proposition \ref{propDGSI}.
   ${\boldsymbol\gamma}$ denotes an arbitrary $k$-Ising matrix.
   The colors of the formulas for the distances match the colors of the corresponding line segments.
   The distances are shown for the example $N=8$ and $k=2$.
   }
  \label{FIGI}
\end{figure}

We will closer examine the structure of Ising states. Let us denote by ${\mathcal I}\subset {\mathcal G}$
the set of Gram matrices with rank one, i.~e., the set of Gram matrices of Ising states, or, shortly, ``Ising matrices".
Such Gram matrices have only entries of the form $\pm 1$.
${\mathcal I}$ has $2^{N-1}$ elements since a total reflection of all spins does not change the Gram matrix.

\begin{lemma}\label{lemmaB}
Every symmetry $\sigma\in {\sf Sym}$ maps ${\mathcal I}$ onto  ${\mathcal I}$.
The special Gram matrix ${\mathbbm 1}_N$ is the barycenter of all Gram matrices of Ising states. It is a fixed point for
all $\sigma\in {\sf Sym}$.
\end{lemma}

\noindent{\bf Proof}\\
It follows from the representation of permutations (\ref{DSG7}) and  partial reflections (\ref{DSG9a}) that symmetries
do not change the rank of Gram matrices. For the second claim of the lemma
consider the barycenter $G=\frac{1}{N}\sum_{{\boldsymbol\gamma}\in{\mathcal I}}\,{\boldsymbol\gamma}$.
Let $1\le\mu<\nu\le N$ and consider the set of Ising spin configurations represented by vectors ${\mathbf s}$ with entries $\pm 1$.
Without loss of generality we may assume ${\mathbf s}_\mu=+1$.
Since the partial reflection $\rho_{\{\nu\}}$ is a bijection of this set there are as much spins with ${\mathbf s}_\nu=+1$
as those with ${\mathbf s}_\nu=-1$. Hence the sum of ${\boldsymbol\gamma}_{\mu\nu}={\mathbf s}_\mu\cdot {\mathbf s}_\nu$
over all ${\boldsymbol\gamma}\in{\mathcal I}$ vanishes. The only non-vanishing matrix elements of $G$ are thus $G_{\mu\mu}=1$
and hence $G={\mathbbm 1}_N$.  The last claim of the lemma follows from the representation of permutations
(\ref{DSG7}) and  partial reflections (\ref{DSG9a}) or, alternatively, from the first statement of the lemma.  \hfill$\Box$\\

Any Ising spin configuration ${\mathbf s}$ can be characterized by the subset ${\mathcal K}\subset \{1,\ldots,N\}$ of indices $\mu$
with ${\mathbf s}_\mu=\downarrow$. Let $A_{\mathcal K}$ denote the corresponding Gram matrix and ${\mathcal I}_k$ the set of
Ising matrices with $\left| {\mathcal K}\right|=k$. ${\boldsymbol\gamma}\in{\mathcal I}_k$ will also be called an
``$k$-Ising matrix".
In the case of $k=0$  or $k=1$ we simply write $A_\mu,\;\mu=0,\ldots,N$
for the corresponding  Ising matrix. $A_0$ is a matrix completely filled with $+1$.
We avoid the ambiguity due to total reflections of Ising states by
restricting the domain of $k$ to $0\le k \le \lfloor \frac{N}{2} \rfloor $. Obviously, $\left| {\mathcal I}_k \right| = {N \choose k}$
for $0\le k < \frac{N}{2}$ and  $\left| {\mathcal I}_k \right| =\frac{1}{2} {N \choose N/2}$ for $k=\frac{N}{2}$ in the case of even $N$.
The latter holds since the $ {N\choose N/2}$ Ising states with $k=N/2$ are pairwise equivalent by means of total reflections.
We have the disjoint decomposition
\begin{equation}\label{DGSI1}
 {\mathcal I}=\bigcup_{k=0,\ldots,\lfloor \frac{N}{2} \rfloor } {\mathcal I}_k
 \;.
\end{equation}
For example, in the case $N=4$ we have  ${\mathcal I}={\mathcal I}_0 \cup {\mathcal I}_1 \cup {\mathcal I}_2 $ corresponding to
$2^3 =8 = 1+4+3$. Since  permutations of Ising spin vector components do not change the number of spins $\downarrow$
the corresponding symmetries will map ${\mathcal I}_k$ onto itself. In contrast, partial reflections will change the number $k$.

Next we consider distances between Ising matrices, i.~e., the norm of its difference.
Here ``distance" and ``norm" will be understood in the sense of the Euclidean scalar product
$\langle A|B\rangle =\mbox{Tr } A\,B$ in the space of real, symmetric $N\times N$-matrices. Since we will often consider differences between Ising matrices and the barycenter ${\mathbbm 1}$
we will introduce the abbreviation  $\hat{\boldsymbol\gamma}\equiv {\boldsymbol\gamma}\,-{\mathbbm 1}$. We state the main results:
\begin{prop}\label{propDGSI}
(i) All Ising matrices  ${\boldsymbol\gamma}\in{\mathcal I}$ satisfy
\begin{equation}\label{DGSI1a}
  || \hat{\boldsymbol\gamma}||=\sqrt{N(N-1)}\;.
\end{equation}

(ii) For all $k$-Ising matrices ${\boldsymbol\gamma}\in{\mathcal I}_k$ there holds
\begin{equation}\label{DGSI2}
\langle  \hat{\boldsymbol\gamma}| \hat{A_0}\rangle = (N-2k)^2-N\;.
\end{equation}

(iii) The distance between $k$-Ising matrices ${\boldsymbol\gamma}\in{\mathcal I}_k$ and $A_0$ is given by
\begin{equation}\label{DGSI3}
 || {\boldsymbol\gamma}-A_0|| = 2\,\sqrt{2\, k\,(N-k)}\,.
\end{equation}

(iv) The barycenter of $k$-Ising matrices $B_k\equiv \frac{1}{|{\mathcal I}_k|}\sum_{{\boldsymbol\gamma}\in{\mathcal I}_k}{\boldsymbol\gamma}$
lies on the line connecting ${\mathbbm 1}$ and $A_0$. More precisely,
\begin{equation}\label{DGSI4}
  B_k =\left( 1-\frac{4 k (N-k)}{N (N-1)}\right)\,A_0+ \frac{4 k (N-k)}{N (N-1)}{\mathbbm 1}
  \;.
\end{equation}
\end{prop}

\noindent{\bf Proof}\\
(i) All Ising matrices have $N^2$ entries $\pm 1$; after subtracting ${\mathbbm 1}$ there remain $N(N-1)$ entries $\pm 1$.
Hence $|| \hat{\boldsymbol\gamma}||^2=\langle\hat{\boldsymbol\gamma}|\hat{\boldsymbol\gamma}\rangle ={N(N-1)},$
from which the claim follows.\\

(ii)  First we consider $\langle {\boldsymbol\gamma}| {A_0}\rangle$ which is the sum of the matrix elements of ${\boldsymbol\gamma}$.
Since it is the same for all ${\boldsymbol\gamma}\in {\mathcal I}_k$ we may choose ${\boldsymbol\gamma}=A_{\mathcal K}$ where
${\mathcal K}=\{1,\ldots,k\}$. Then  $A_{\mathcal K}$ consists of two rectangles of size $k^2$ and $(N-k)^2$ filled with $+1$
and two rectangles of size $k\times(N-k)$ filled with $-1$. This yields\\
 $\langle A_{\mathcal K}| A_0\rangle=k^2+(N-k)^2 -2 k(N-k)=(N-2k)^2$. Subtracting ${\mathbbm 1}$ from both arguments results
 in filling the diagonal $A_{\mathcal K}$ with zeroes, which gives a contribution of $-N$. Summarizing,
 $\langle  \hat{\boldsymbol\gamma}| \hat{A_0}\rangle = (N-2k)^2-N$.\\
(iii) This follows from
\begin{eqnarray}
\label{DGSI4a}
 || {\boldsymbol\gamma}-A_0||^2&=&|| \hat{\boldsymbol\gamma}-\hat{A_0}||^2 \\
 \label{DGSI4b}
   &=& \langle \hat{\boldsymbol\gamma}-\hat{A}_0 | \hat{\boldsymbol\gamma}-\hat{A_0}\rangle \\
   \label{DGSI4c}
  &=& || \hat{\boldsymbol\gamma}||^2+ ||\hat{A}_0||^2-2\,\langle  \hat{\boldsymbol\gamma}| \hat{A_0}\rangle\\
  \nonumber
   &\stackrel{(\ref{DGSI1a})(\ref{DGSI2})}{=}&2N(N-1)-2 ((N-2k)^2-N)\\
   \label{DGSI4d}
   &=&8 k (N-k)\;.
\end{eqnarray}

(iv) Let $1\le \mu<\nu\le N$ and consider $\left(B_k\right)_{\mu\nu}$. The subgroup ${\sf Per}$ of symmetries induced by
permutations $\pi\in S_N$ leaves ${\mathcal I}_k$ and hence $B_k$ invariant. Moreover, $S_N$ operates transitively on the set of all
subsets of $\{1,\ldots,N\}$ with two elements. This implies that $\left(B_k\right)_{\mu\nu}$ is the same for all
$1\le \mu<\nu\le N$. Hence $B_k$ is of the form ${\mathbbm 1}+\alpha\,\hat{A_0},\;\alpha\in{\mathbbm R}$ (recall that $\hat{A_0}$ is a matrix
completely filled with $+1$ except the zero diagonal). To determine $\alpha$ recall that by (\ref{DGSI2}) all  $\hat{\boldsymbol\gamma}\in{\mathcal I}_k$
have the same orthogonal  projection  onto the line connecting ${\mathbbm 1}$ and $A_0$. Hence also $\hat{B_k}$ has the
same projection and therefore
\begin{eqnarray}\label{DGSI5a}
  (N-2k)^2-N&=&\langle \hat{B_k} | \hat{A_0}\rangle =\langle \alpha\hat{A_0} | \hat{A_0}\rangle  \\
  \label{DGSI5b}
  &=&\alpha ||\hat{A_0}||^2\stackrel{(\ref{DGSI1a})}{=} \alpha\,N(N-1)\,,\\
  \label{DGSI5c}
  \alpha  &=& \frac{(N-2k)^2-N}{N(N-1)} \\
  \label{DGSI5d}
   &=&1-\frac{4k(N-k)}{N(N-1)}
   \;.
\end{eqnarray}
It follows that $B_k= {\mathbbm 1}+\alpha\,(A_0-{\mathbbm 1})=\alpha\,A_0+(1-\alpha){\mathbbm 1}$
which, together with (\ref{DGSI5d}), implies (\ref{DGSI4}).   \hfill$\Box$\\

Although it is difficult, if not impossible, to visualize convex sets of dimension ${N \choose 2}$ for N>3, the preceding proposition
provides a certain insight into the geometry of ${\mathcal G}$ by means of its ``skeleton" ${\mathcal I}$ of Ising matrices, see Figure \ref{FIGI}.
Ising matrices are prominent extremal points of ${\mathcal G}$ lying at a sphere of radius $\sqrt{N(N-1)}$ with center ${\mathbbm 1}$.
Viewed from the Ising matrix $A_0$ there are different ``shells" ${\mathcal I}_k,\; k=1,\ldots, \lfloor N/2 \rfloor$ of $k$-Ising
matrices with constant distances (\ref{DGSI3}) from $A_0$ and constant orthogonal projections (\ref{DGSI2}) onto the line
between ${\mathbbm 1}$ and $A_0$. There is nothing special about the Ising matrix $A_0$; since the subgroup ${\sf Rf}$ of symmetries
generated by partial reflections $\rho_{\mathcal T}$ operates transitively on ${\mathcal I}$ the analogous geometrical picture
of concentric shells arises with every ${\boldsymbol\gamma}\in{\mathcal I}$ as the center. This situation remotely resembles
the geometry of, say, the icosahedron, where each vertex is surrounded by concentric shells of other vertices.\\

Both approaches, the Lagrange variety approach and the Gram set approach, will be combined in order to investigate the general spin triangle.
This case will be interesting in its own right as well as to illustrate the concepts introduced so far.

\section{Ground states of the classical spin triangle}\label{sec:ST}
\begin{figure}[h]
\vspace{5cm}
  \centering
    \includegraphics[width=0.3\linewidth,viewport =40mm 0mm 80mm 80mm]{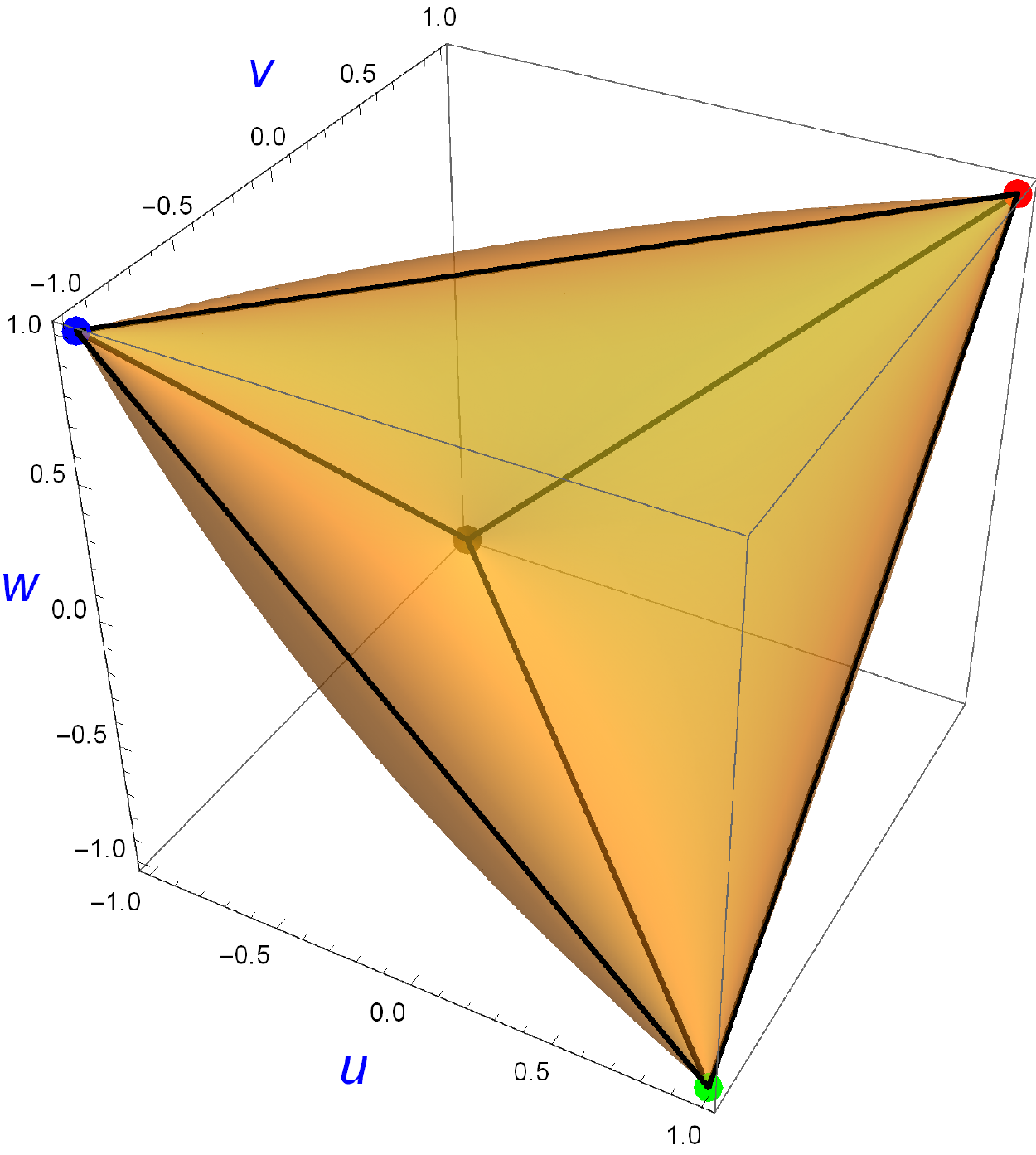}
  \caption[The minimal eigenvalue]
  {Plot of the convex Gram set ${\mathcal G}$ defined in (\ref{ST4}) for $N=3$ together with four extremal points corresponding to Ising states
  and the six faces generated by them.
  The points of ${\mathcal G}$ represent equivalence classes of spin configurations w.~r.~t.~uniform rotations and/or reflections.
  Obviously, ${\mathcal G}$  has the full tetrahedral symmetry.}
  \label{FST2}
\end{figure}

In the case $N=3$ the ground states of the general Heisenberg Hamiltonian
\begin{equation}\label{ST1}
  H({\mathbf s})=J_1\, {\mathbf s}_2\cdot{\mathbf s}_3+J_2\, {\mathbf s}_3\cdot{\mathbf s}_1+J_3\, {\mathbf s}_1\cdot{\mathbf s}_2
\end{equation}
can be completely determined. It is clear that a common positive factor in (\ref{ST1}) will be irrelevant for the ground states,
hence it would suffice to restrict the vector  ${\mathbf J}\equiv\left(J_1,J_2,J_3\right)$ to the unit sphere in ${\mathbb R}^3$;
but we will not make use of this simplification in what follows.\\
The Gram matrices $G\in{\mathcal G}_3$ have the form
\begin{equation}\label{ST2}
 G=\left(
\begin{array}{ccc}
1& u&v\\
u&1&w\\
v&w&1
\end{array}
\right)
\equiv[u,v,w]
\;,
\end{equation}
such that the Hamiltonian (\ref{ST1}) assumes the form
\begin{equation}\label{ST2a}
H(u,v,w)=J_1\,w+J_2\,v+J_3\,u\;,
\end{equation}
and
\begin{equation}\label{ST3}
\det\,G = \det\,[u,v,w] =1+2 u v w -(u^2+v^2+w^2)
\;.
\end{equation}
The Gram set, already defined in \cite{SL03}, assumes the form
\begin{equation}\label{ST4}
{\mathcal G}=\left\{\left. [u,v,w]\right|
u^2,v^2,w^2\le 1,\,\det[u,v,w]\ge 0
\right\}
\;.
\end{equation}
It is $3$-dimensional and hence can be visualized, see Figure \ref{FST2}. All possible ground states lie at the boundary of ${\mathcal G}$.
This boundary consists of all Gram matrices $G$ with $\mbox{rank}(G)\le 2$ hence the ground states are $1$- or $2$-dimensional
(Ising states or co-planar states). They can be divided into three classes:
\begin{enumerate}
  \item[I] The four special extremal points\\
  \begin{eqnarray}\label{FST2a}
    A_0 &=&[1,1,1], \\
    \label{FST2b}
    A_1 &=& [-1,-1,1], \\
    \label{FST2c}
    A_2 &=& [-1,1,-1], \\
    \label{FST2d}
    A_3 &=&[1,-1,-1],
  \end{eqnarray}
  corresponding to the Ising matrices considered in section \ref{sec:DGSI}.
  They correspond to the $2^{N-1}=4$ different Ising states due to rotational and/or reflectional equivalence.
  \item[II] A $2$-dimensional manifold of extremal points defined by
  \begin{eqnarray}\nonumber
  M&=&\{[u,v,w]\left|-1<u,v,w<1,\,\det[u,v,w]=0\right.\}.\\
  \label{FST2e}
  \end{eqnarray}
  The corresponding ground states are co-planar.
  \item[III]Six open intervals joining all pairs of extremal points of type I:
  \begin{eqnarray}
  \label{FST2f}
    \overline{A_0,A_3} &=& \{\,[1,x,x]\,|\,-1<x<1\,\}, \\    \label{FST2g}
     \overline{A_0,A_2} &=& \{\,[x,1,x]\,|\,-1<x<1\,\}, \\   \label{FST2h}
     \overline{A_0,A_1} &=& \{\,[x,x,1]\,|\,-1<x<1\,\}, \\   \label{FST2i}
      \overline{A_2,A_3} &=& \{\,[x,-x,-1]\,|\,-1<x<1\,\} ,\\    \label{FST2j}
       \overline{A_1,A_3} &=& \{\,[x,-1,-x]\,|\,-1<x<1\,\}, \\   \label{FST2k}
        \overline{A_1,A_2} &=& \{\,[-1,x,-x]\,|\,-1<x<1\,\}
        \;.
     \end{eqnarray}
     The corresponding ground states are the special co-planar states such that two spins are parallel or anti-parallel.
\end{enumerate}

The closed faces of ${\mathcal G}$, except the trivial ones $\emptyset$ and ${\mathcal G}$, are the extremal points of the classes I and II and the six faces obtained by adding the end points to the six open intervals of class III. The manifold of class II consists of four connected parts divided  by the six intervals of class III.
Interestingly, these intervals fit smoothly into the manifold such that the union of the points of classes II and III again forms a manifold.
This can be seen by simultaneously solving the equations $\det\,[u,v,w]=0$ and $\nabla\det\,[u,v,w]={\mathbf 0}$. The only solutions
are the four extremal points $A_0,A_1,A_2,A_3$ of class I. This implies that the variety defined by $\det\,[u,v,w]=0$,
if restricted to ${\mathcal G}$, becomes a smooth two-dimensional
manifold if the four points $A_0,A_1,A_2,A_3$ are excluded.\\
Recall from section \ref{sec:DGS} that the group ${\sf Sym}$ of symmetries of ${\mathcal G}$ is of order $N!\,2^{N-1}$ which equals
$24$ for $N=3$. Obviously, for the spin triangle ${\sf Sym}$ can be identified with the tetrahedral group $T_h$ including reflections.
Each transformation $\gamma\in{\sf Sym}$ induces a permutation of the four extremal points $A_0,A_1,A_2,A_3$; and vice versa, every such permutation
is induced by a unique symmetry $\gamma\in{\sf Sym}$. This illustrates the well-known isomorphism $T_h\cong S_4$. The use of symmetries
simplifies the following investigations of ground states.\\
As remarked above the variety $\det\,[u,v,w]=0$ has exactly the four singular points $A_0,A_1,A_2,A_3$. Due to the symmetries of ${\mathcal G}$
it suffices to consider one of these singular points, say, $A_0=[1,1,1]$ corresponding to the Ising state $\uparrow\uparrow\uparrow$.
In the infinitesimal neighbourhood of $A_0$ the variety $\det\,[u,v,w]=0$ assumes the form of a circular cone $C_0$. This can be seen, e.~g.~,
by expanding $\det\,[u,v,w]$ at  $A_0$ up to second order. The set of affine functionals ${\mathcal J}:{\mathcal G}\longrightarrow{\mathbb R}$
defined by ${\mathcal J}(G)\equiv\mbox{Tr}(G\,{\mathbb J})=(u,v,w)\cdot {\mathbf J}$ that assume their minimal value at  $A_0$ can be identified
with the dual cone $C_0'$ of $C_0$. Physically, this dual cone corresponds to the set of ``ferromagnetic Hamiltonians" that have ground states
of the form $\uparrow\uparrow\uparrow$. In terms of the vectors ${\mathbf J}=(J_1,J_2,J_3)$ this ``ferromagnetic cone" $C_0'$ can be
characterized by
\begin{equation}\label{ST4a}
 {\mathbf J}\cdot(1,1,1)=J_1+J_2+J_3 \le -||{\mathbf J}||
 \;.
\end{equation}
Note that $C_0'$ is
larger than the octant defined by $J_1,J_2,J_3\le 0$ , where the ferromagnetic nature of the Hamiltonian is self-evident.\\
According to the tetrahedral symmetry of ${\mathcal G}$ these results can be immediately transferred to the other extremal points
$A_1,A_2,A_3$. For example, if ${\mathbf E}_i$ denote the four vectors with the coordinates $(w,v,u)$ of the extremal points $A_i,\,i=0,1,2,3$,
then the equations for the corresponding cones $C_i'$ that generalize (\ref{ST4a}) read:
\begin{equation}\label{ST5}
 {\mathbf J}\cdot{\mathbf E}_i \le -||{\mathbf J}||,\; i=0,1,2,3
  \;.
\end{equation}
If ${\mathbf J}$ does not satisfy either of the four conditions (\ref{ST5}) then its ground states will be co-planar and correspond to a point
$[u,v,w]$ of the manifold $M$ of class II considered above. The map from ground states to Hamiltonians
$[u,v,w]\mapsto (J_1,J_2,J_3)$ can be easily determined:
${\mathbf J}$ can be chosen  proportional to the gradient of ${\mathcal D}\equiv\det[u,v,w]$ w.~r.~t.~the three variables $u,v,w$:
\begin{eqnarray}\label{ST6a}
J_1&=&\mu \frac{\partial{\mathcal D}}{\partial w}=2\mu(u v-w),\\
\label{ST6b}
J_2&=&\mu\frac{\partial{\mathcal D}}{\partial v}=2\mu(w u-v),\\
\label{ST6c}
J_3&=&\mu\frac{\partial{\mathcal D}}{\partial u}=2\mu(v w -u)\;.
\end{eqnarray}
The converse is also straightforward: The solution of the system of equations (\ref{ST6a}),(\ref{ST6b}),(\ref{ST6c}) together with ${\mathcal D}=0$
for the unknowns $u,v,w,\mu$ reads:
\begin{eqnarray}\label{ST7a}
u&=& \frac{{J_1}}{2 {J_2}} \left(\frac{{J_2}^2}{{J_3}^2}-1\right)-\frac{{J_2}}{2 {J_1}},\\
\label{ST7b}
v&=& \frac{{J_1}}{2  {J_3}} \left(\frac{{J_3}^2}{{J_2}^2}-1\right)-\frac{{J_3}}{2{J_1}},\\
\label{ST7c}
w&=&\frac{{J_2}}{2   {J_3}} \left(\frac{{J_3}^2}{{J_1}^2}-1\right)-\frac{{J_3}}{2 {J_2}},\\
\label{ST7d}
\mu&=&
\scriptstyle
\frac{-2{J_1}^3 {J_2}^3 {J_3}^3}{{J_1}^4
   \left({J_2}^2-{J_3}^2\right)^2-2 {J_1}^2{J_2}^2 {J_3}^2
   \left({J_2}^2+{J_3}^2\right)+{J_2}^4 {J_3}^4}\;.
\end{eqnarray}

We will provide an example. Assume a Hamiltonian with $J_1=1,\,J_2=-2,\,J_3=-1$. One checks that the four conditions (\ref{ST5})
are not satisfied and hence this Hamiltonian does not have an Ising ground state. Inserting the coupling constants into (\ref{ST7a}) -- (\ref{ST7d})
yields $u=\frac{1}{4},\,v=\frac{7}{8},\,w=-\frac{1}{4},\,\mu =\frac{16}{15}$ and a ground state energy of
$E_{min}=J_1\,w+J_2\,v+J_2\,u=-\frac{9}{4}$.
A possible ground state is hence
${\mathbf s}_1=(1,0),\,{\mathbf s}_2=(u,\sqrt{1-u^2})=\left(\frac{1}{4},\frac{\sqrt{15}}{4}\right),\,
{\mathbf s}_3= (v,-\sqrt{1-v^2})=\left(\frac{7}{8},-\frac{\sqrt{15}}{8}\right)$. Note that the choice ${\mathbf s}_3= (v,+\sqrt{1-v^2})$ would not give the correct value of ${\mathbf s}_2\cdot{\mathbf s}_3 = w =-\frac{1}{4}$. We will also check the SSE (\ref{D4}):
\begin{eqnarray}\label{STSSEa}
 {\mathbbm J}\,{\mathbf s}&=&\left(
\begin{array}{ccc}
 0 & -\frac{1}{2} & -1 \\
 -\frac{1}{2} & 0 & \frac{1}{2} \\
 -1 & \frac{1}{2} & 0 \\
\end{array}
\right)
\left(
\begin{array}{cc}
 1 & 0 \\
 \frac{1}{4} & \frac{\sqrt{15}}{4} \\
 \frac{7}{8} & -\frac{\sqrt{15}}{8} \\
\end{array}
\right)\\
\label{STSSEb}
&=&
\left(
\begin{array}{cc}
 -1 & 0 \\
 -\frac{1}{16} & -\frac{\sqrt{15}}{16} \\
 -\frac{7}{8} & \frac{\sqrt{15}}{8} \\
\end{array}
\right)\\
\label{STSSEc}
&=&
\left(
\begin{array}{ccc}
 -1 & 0 & 0 \\
 0 & -\frac{1}{4} & 0 \\
 0 & 0 & -1 \\
\end{array}
\right)
\left(
\begin{array}{cc}
 1 & 0 \\
 \frac{1}{4} & \frac{\sqrt{15}}{4} \\
 \frac{7}{8} & -\frac{\sqrt{15}}{8} \\
\end{array}
\right).
\end{eqnarray}
The negative mean value of the Lagrange parameters $-\overline{\kappa}=-\frac{3}{4}$ is also the lowest eigenvalue $j_{min}$
of the dressed ${\mathbbm J}$-matrix in the ground state gauge
\begin{equation}\label{STSSEd}
 {\mathbbm J}({\boldsymbol\lambda})=
 \left(
\begin{array}{ccc}
 \frac{1}{4} & -\frac{1}{2} & -1 \\
 -\frac{1}{2} & -\frac{1}{2} & \frac{1}{2} \\
 -1 & \frac{1}{2} & \frac{1}{4} \\
\end{array}
\right)
\end{equation}
with two-fold degeneracy. This complies with $E_{min}=N\,j_{min}= 3\,(-\frac{3}{4})=-\frac{9}{4}$.

\vspace{3cm}
\begin{figure}[ht]
  \centering
    \includegraphics[width=0.25\linewidth,viewport =40mm 0mm 80mm 80mm]{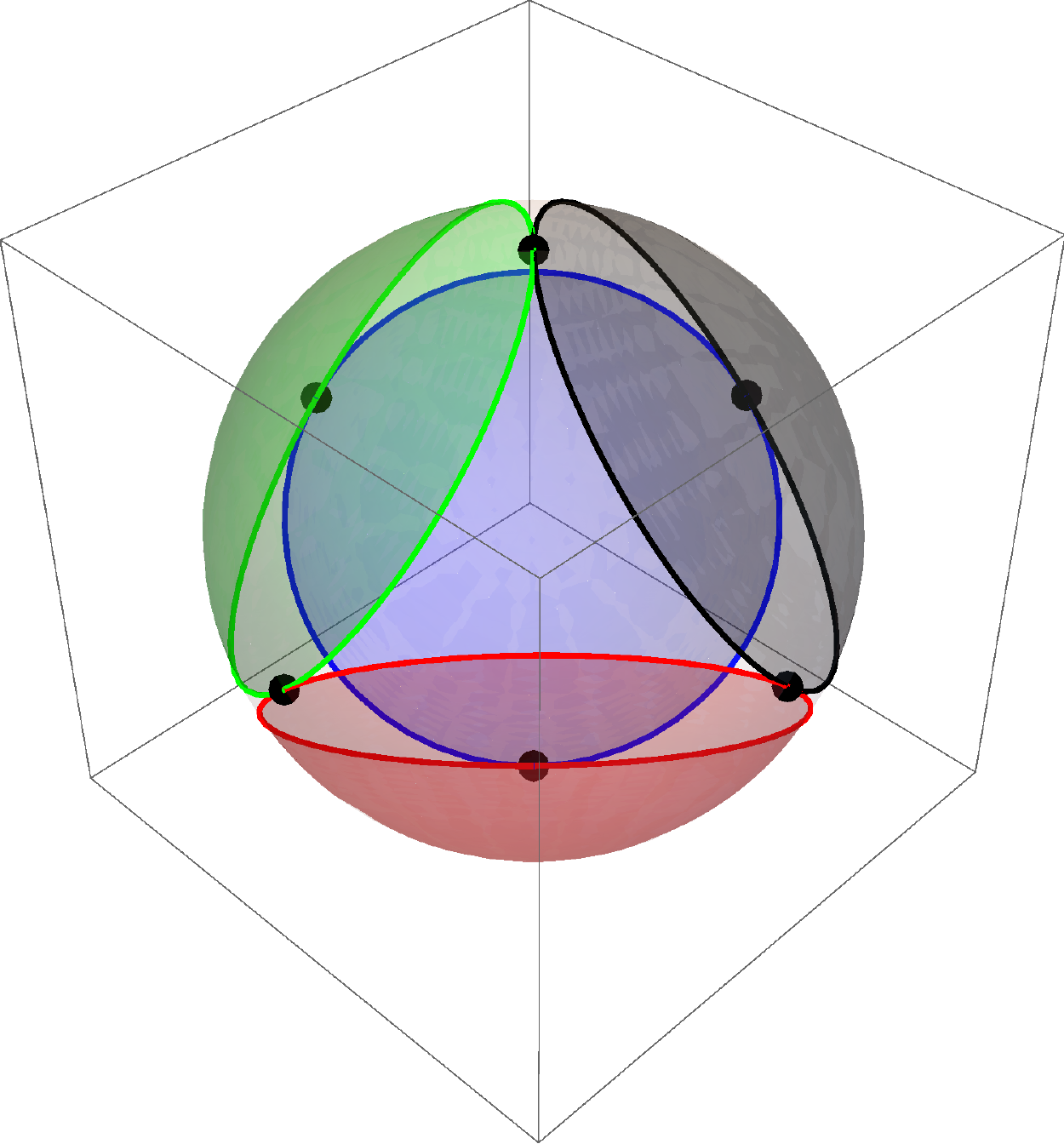}
  \caption[The minimal eigenvalue]
   {The phase diagram of the set of spin coupling coefficients ${\mathbf J}=(J_1,J_2,J_3)$ restricted to the unit sphere. The four colored
   spherical shells correspond to systems that have Ising ground states of the kind $A_i,\;i=0,1,2,3$. The six common points (black dots) correspond
   to degenerate ground states (\ref{FST2f}) to  (\ref{FST2k}). The area between the spherical shells corresponds to systems with co-planar
   ground states parametrized by the points of the manifold $M$ defined  in (\ref{FST2e}).}
  \label{FST6}
\end{figure}

Summarizing, we can divide the unit sphere of the ${\mathbf J}$-space into various parts, see figure \ref{FST6}. First, we have four closed spherical shells
${\mathcal Z}_i$ that span the cones $C_i'$ considered above and belong to the four Ising ground states
$A_i,\;i=0,1,2,3$. It turns out that each shell has three common points with the other three shells. The resulting number of $6$
points in ${\mathbf J}$-space corresponds to the $6$ degenerate ground states of class III. The remaining part of the unit sphere
not belonging to any  ${\mathcal Z}_i$ represents all normal directions of the manifold $M$ of class II. Together with $M$
it has four connected components.\\

\begin{figure}[ht]
  \centering
    \includegraphics[width=0.3\linewidth,viewport =40mm 0mm 80mm 80mm]{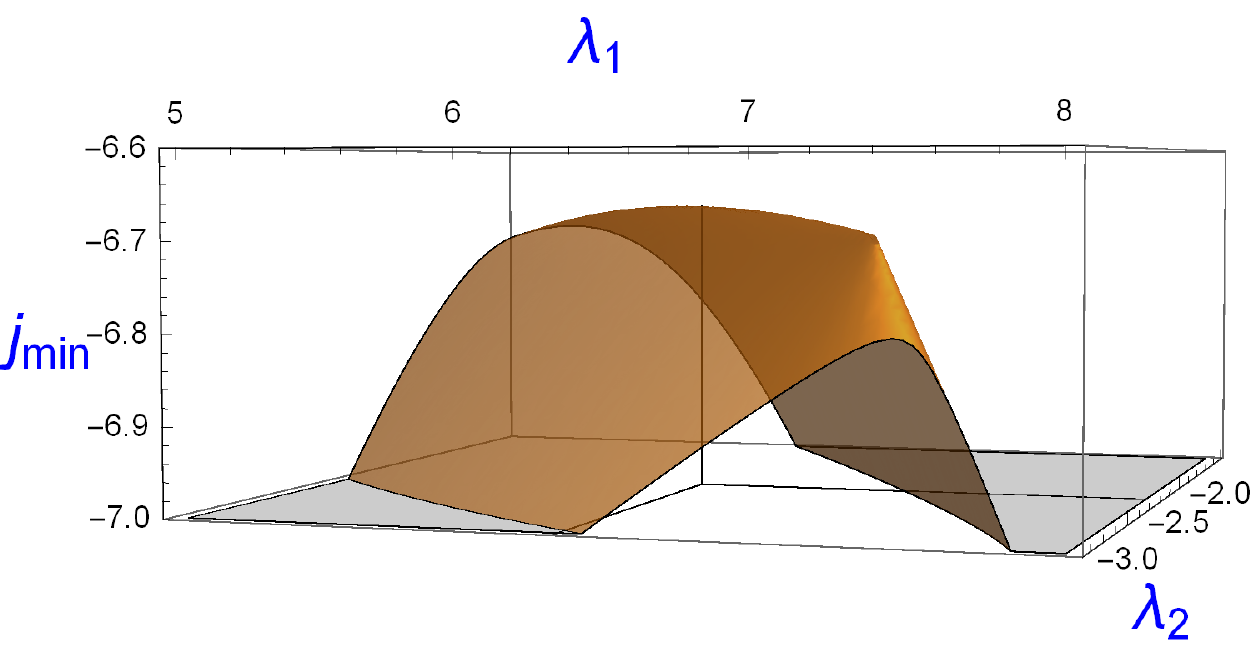}
  \caption[The minimal eigenvalue]
  {The minimal eigenvalue $j_{min}(\lambda_1,\lambda_2)$ of ${\mathbb J}(\lambda_1,\lambda_2)$, where $J_1=3,\,J_2=-6,\,J_3=-7$.
  At the point $\lambda_1=\frac{19}{3},\,\lambda_2=-\frac{8}{3}$ the function  $j_{min}$ has a smooth maximum
  assuming the value $\widehat{\jmath}_{min}=-\frac{20}{3}$ corresponding to an Ising ground state $\uparrow\uparrow\uparrow$.
  }
  \label{FST4}
\end{figure}

In the following we consider four typical positions of ${\mathbf J}$ and the corresponding types of ground states.
Let ${\mathbf J}=(6,-12,-14)$; since $J_1+J_2+J_3=-20<-\sqrt{376}=-||{\mathbf J}||$, it follows that
 ${\mathbf J}$ lies in the interior of the ferromagnetic cone. W.~r.~t.~the Ising state $\uparrow\uparrow\uparrow$ the ground state
 gauge ${\mathbb J}({\boldsymbol\lambda})$ assumes the form
 \begin{equation}\label{ST9}
{\mathbb J}({\boldsymbol\lambda})= \left(
\begin{array}{ccc}
 \frac{19}{3} & -7 & -6 \\
 -7 & -\frac{8}{3} & 3 \\
 -6 & 3 & -\frac{11}{3} \\
\end{array}
\right)
 \;,
\end{equation}
 and its lowest eigenvalue $\hat{\jmath}_{min}=-\frac{20}{3}$ is non-degenerate with eigenvector $(1,1,1)$.
Accordingly, at the point ${\boldsymbol\lambda}=(\frac{19}{3},-\frac{8}{3})$ the function $j_{min}({\boldsymbol\lambda})$ has a smooth
maximum, see Figure \ref{FST4}. \\

If we increase $J_1$ of the preceding example from the value $6$ to $\frac{84}{13}=6.\overline{461538}$, ${\mathbf J}$
moves to the boundary of the ferromagnetic cone
since $J_1+J_2+J_3=-\frac{254}{13}=-||{\mathbf J}||$. The plane perpendicular to ${\mathbf J}$ will now be
tangent to the manifold $M$ of class II
at the extremal point $A_0$. The ground state
 gauge w.~r.~t.~$\uparrow\uparrow\uparrow$ assumes the form

\vspace{3cm}
\begin{figure}[ht]
  \centering
    \includegraphics[width=0.3\linewidth,viewport =40mm 0mm 80mm 80mm]{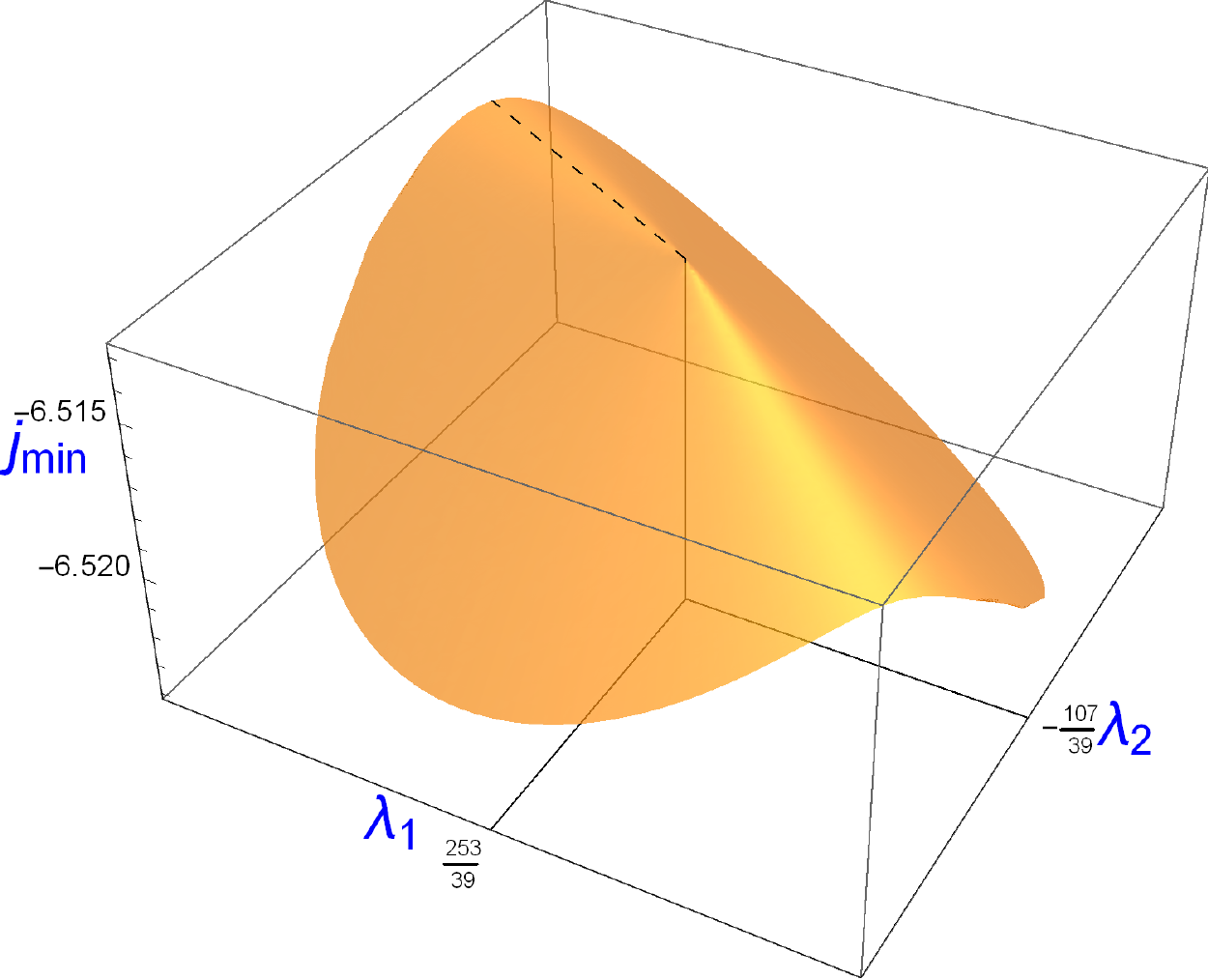}
  \caption[The minimal eigenvalue]
  {The minimal eigenvalue $j_{min}(\lambda_1,\lambda_2)$ of ${\mathbb J}(\lambda_1,\lambda_2)$, where $J_1=-7,\,J_2=-6,\,J_3=\frac{42}{13}$.
  At the point $\lambda_1=\frac{253}{39},\,\lambda_2=-\frac{107}{39}$ the function  $j_{min}$  assumes its maximal value $\widehat{\jmath}_{min}=-\frac{254}{39}$.
  At this point the function has a tangent cone
 with a horizontal direction (dashed line) at the angle $\varphi=\pi-\arctan\frac{7}{13}$.}
  \label{FST3}
\end{figure}

 \begin{equation}\label{ST10}
{\mathbb J}({\boldsymbol\lambda})= \left(
\begin{array}{ccc}
 \frac{253}{39} & -7 & -6 \\
 -7 & -\frac{107}{39} & \frac{42}{13} \\
 -6 & \frac{42}{13} & -\frac{146}{39} \\
\end{array}
\right)
 \;,
 \end{equation}
and its lowest eigenvalue $\hat{\jmath}_{min}=-\frac{254}{39}$ becomes two-fold degenerate. The corresponding eigenspace
spanned by $(6/13, 0, 1)$ and $(7/13, 1, 0)$ is still elliptic
since the solution $\Delta$ of the system of equations corresponding to (\ref{DSG10}) has the eigenvalues $2,0$.
But only one eigenvector exists that gives rise to a spin configuration, namely the sum of the above two eigenvectors
which again corresponds to $\uparrow\uparrow\uparrow$. This is the example where the eigenspace of ${\mathbb J}({\boldsymbol\lambda})$ is elliptic but not completely elliptic, that was announced on p.~2. At its maximum the function $j_{min}(\lambda_1,\lambda_2)$
is not smooth but has a tangent cone with a horizontal direction, see figure \ref{FST3}.\\

\vspace{2cm}
\begin{figure}[ht]
  \centering
    \includegraphics[width=0.3\linewidth,viewport =40mm 0mm 80mm 80mm]{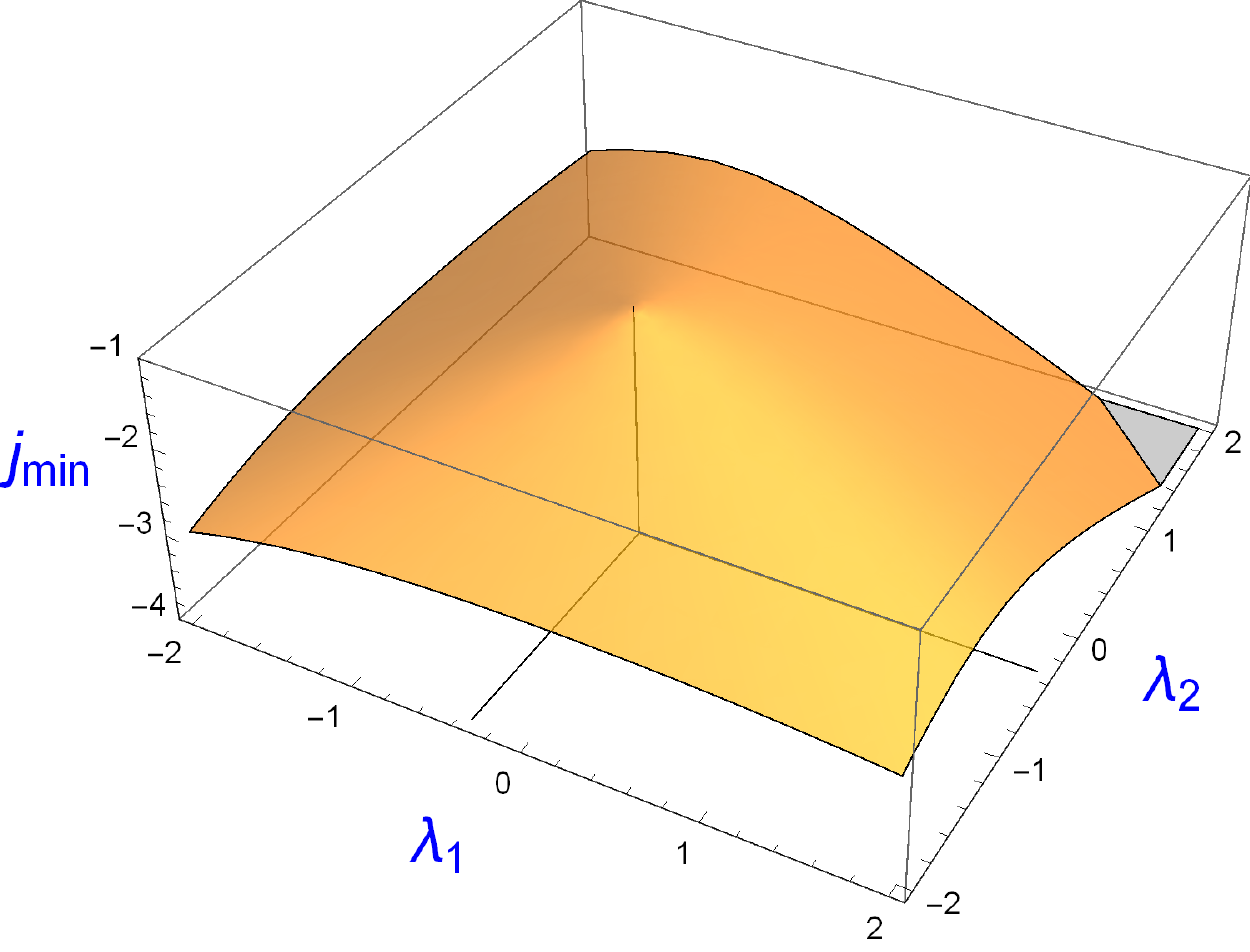}
  \caption[The minimal eigenvalue]
   {The minimal eigenvalue $j_{min}(\lambda_1,\lambda_2)$ of ${\mathbb J}(\lambda_1,\lambda_2)$, where $J_1=1,\,J_2=J_3=-\sqrt{2}$.
  At the point $\lambda_1=\lambda_2=-1/3$ the function  $j_{min}$ has a maximum with the value $\widehat{\jmath}_{min}=-4/3 $ and its graph assumes the form of
  a circular cone in the infinitesimal neighbourhood of the maximum.}
  \label{FST1}
\end{figure}

Next we choose an example where ${\mathbf J}$ assumes its minimum at an extremal point belonging to the manifold $M$ of class II, see above,
namely $J_1=J_2=-2\sqrt{2},\,J_3=2$. The ground state(s) can be determined by using (\ref{ST7a}) -- (\ref{ST7c}) which yields
$u=0,\,v=w=\frac{1}{\sqrt{2}}$ and a ground state energy $E_{min}=-4$. The corresponding Gram matrix $[u,v,w]$ is, e.~g.~, realized by the spin configuration
${\mathbf s}_1=(1,0),\;{\mathbf s}_2=(0,1),\;{\mathbf s}_3=(\frac{1}{\sqrt{2}},\frac{1}{\sqrt{2}})$.
W.~r.~t.~this configuration the dressed ${\mathbb J}$ matrix in the ground state gauge is obtained as:
\begin{equation}\label{ST11}
 {\mathbb J}({\boldsymbol\lambda})=\left(
\begin{array}{ccc}
 -\frac{1}{3} & 1 & -\sqrt{2} \\
 1 & -\frac{1}{3} & -\sqrt{2} \\
 -\sqrt{2} & -\sqrt{2} & \frac{2}{3} \\
\end{array}
\right).
\end{equation}
It has a doubly degenerate lowest eigenvalue ${\hat{\jmath}}_{min}=-\frac{4}{3}=\frac{1}{3}E_{min}$ with an eigenspace spanned by the vectors ${\mathbf e}_1=(\sqrt{2},0,1)$
and ${\mathbf e}_2=(-1,1,0)$.
At its maximum the function $j_{min}(\lambda_1,\lambda_2)$ is not smooth but has a circular tangent cone, see figure \ref{FST1}.\\

\begin{figure}[ht]
  \centering
    \includegraphics[width=0.35\linewidth,viewport =40mm 0mm 80mm 80mm]{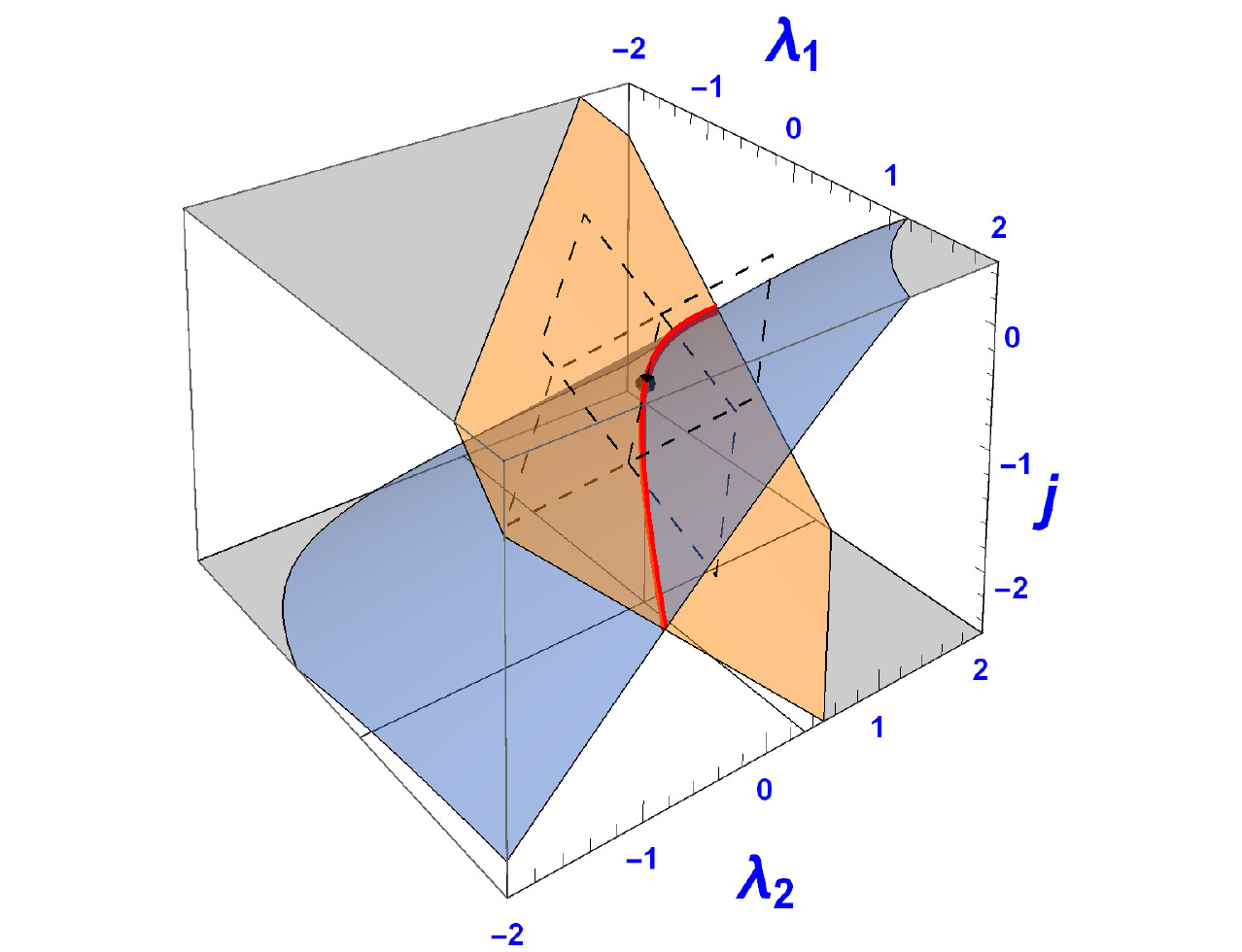}
  \caption[The minimal eigenvalue]
  {The minimal eigenvalue $j_{min}(\lambda_1,\lambda_2)$ of ${\mathbb J}(\lambda_1,\lambda_2)$, where $J_1=J_2=0,\,J_3=1$.
   $j_{min}(\lambda_1,\lambda_2)$ is the minimum of two smooth eigenvalue functions $j_{1}(\lambda_1,\lambda_2)$ and $j_{2}(\lambda_1,\lambda_2)$.
  The intersection between the graphs of $j_{1}(\lambda_1,\lambda_2)$ and $j_{2}(\lambda_1,\lambda_2)$ is shown as a red curve.
  At the point $\lambda_1=\lambda_2=\frac{1}{3}$ the function  $j_{min}$  assumes its maximal value $\widehat{\jmath}_{min}=-\frac{2}{3}$ (black dot).
  At this point the tangent cone degenerates into a wedge generated by the two tangent planes at  $j_{1}(\lambda_1,\lambda_2)$
  and $j_{2}(\lambda_1,\lambda_2)$  with a horizontal intersection (dashed lines).}
  \label{FST5}
  \end{figure}

Finally we remark that the six cases of ${\mathbf J}=(\pm 1,0,0)$ and permutations correspond to disjoint systems where two spins are
(anti)\-ferromagnetically coupled and the third spin is uncoupled. The ground state is hence a degenerate
configuration where two spins are (anti)\-parallel
and the third spin is arbitrary. These are the six cases (\ref{FST2f}) to  (\ref{FST2k}).
As an example we show the graph of $j_{min}(\lambda_1,\lambda_2)$ for the case of ${\mathbf J}=(0,0,1)$, see Figure \ref{FST5}.
At the maximum of $j_{min}(\lambda_1,\lambda_2)$ the elliptic tangent cone degenerates into a wedge.
\\

\vspace{0cm}

\section{Summary and Outlook}\label{sec:SO}
In this paper we have started the investigation of the convex set ${\mathcal G}$ of Gram matrices of $N$ spins. This leads to a
slightly different view on the ground state problem: The points of ${\mathcal G}$ are $O(M)$-equivalence classes of spin configurations
and each Heisenberg Hamiltonian is an affine functional defined on ${\mathcal G}$ such that its minimal energy value is assumed
at a face of ${\mathcal G}$. If the face is an extremal point, the ground state is essentially unique; but if it consists of
more than one point (or, equivalently, has a dimension $>0$) we encounter the effect of additional degeneracy that is familiar from
many examples. Hence we are lead to the general problem to classifying and characterizing 
the faces of ${\mathcal G}$, a problem that we have to defer to subsequent papers. 
An equivalent problem would be the classification of completely elliptic subspaces of ${\mathbbm R}^N$.
In the present paper we have found some partial results on the subset of extremal points corresponding to Ising states
and calculated the complete symmetry group of ${\mathcal G}$. The latter is also of avail for practical purposes.

The second part of the paper was devoted to the ground state problem of the general spin triangle, 
where ${\mathcal G}$ can be visualized as a subset of
${\mathbbm R}^3$ and its structure can be completely clarified. 
It was also possible in this case to visualize  for typical cases the Lagrange variety in the neighborhood of vertical points corresponding to
ground state(s) of suitable Hamiltonians.
As we have shown, there exists an algorithm to calculate the ground state(s)
for any given Heisenberg Hamiltonian, but it is pretty obvious that this cannot be generalized to larger $N$.

\appendix
\section{Elliptic subspaces}\label{sec:ES}
Recall that a ``completely elliptic subspace" (CES) $S$ of ${\mathbbm R}^N$ is defined by the existence of an
$N\times M$-matrix  $W$ such that its $M$ columns form a basis of ${ S}$ and its $N$ rows are unit vectors.
Alternatively, a CES ${S}$ is characterized by the property that for {\it any} $ N\times M$-matrix  $W$
such that its $M$ columns form a basis of ${ S}$ the $N$ rows will lie an a non-degenerate $N-$dimensional ellipsoid.
It will be instructive to determine the completely elliptic subspaces for simple examples.
As an obvious general result we note that permutations of coordinates and partial reflections  $\rho_{\mathcal T}$, see(\ref{DSG8}),
map the class of CES onto itself. In this paper we have determined the CES of  ${\mathbbm R}^3$.

\subsection{Complete elliptic subspaces of ${\mathbbm R}^3$}\label{sec:ES3}
Obviously, there exist exactly four one-dimensional CES $L_\mu,\mu=0,\ldots,3$ of ${\mathbbm R}^3$ spanned by the vectors
$(1,1,1),\;(-1,1,1),\;(1,-1,1),\;(1,1,-1)$ , resp.~. The $L_\mu$ are connected by partial reflections, see above,
and can be visualized by the four space diagonals of the cube.

\vspace{40mm}
\begin{figure}[ht]
  \centering
    \includegraphics[width=0.3\linewidth,viewport =40mm 0mm 80mm 80mm]{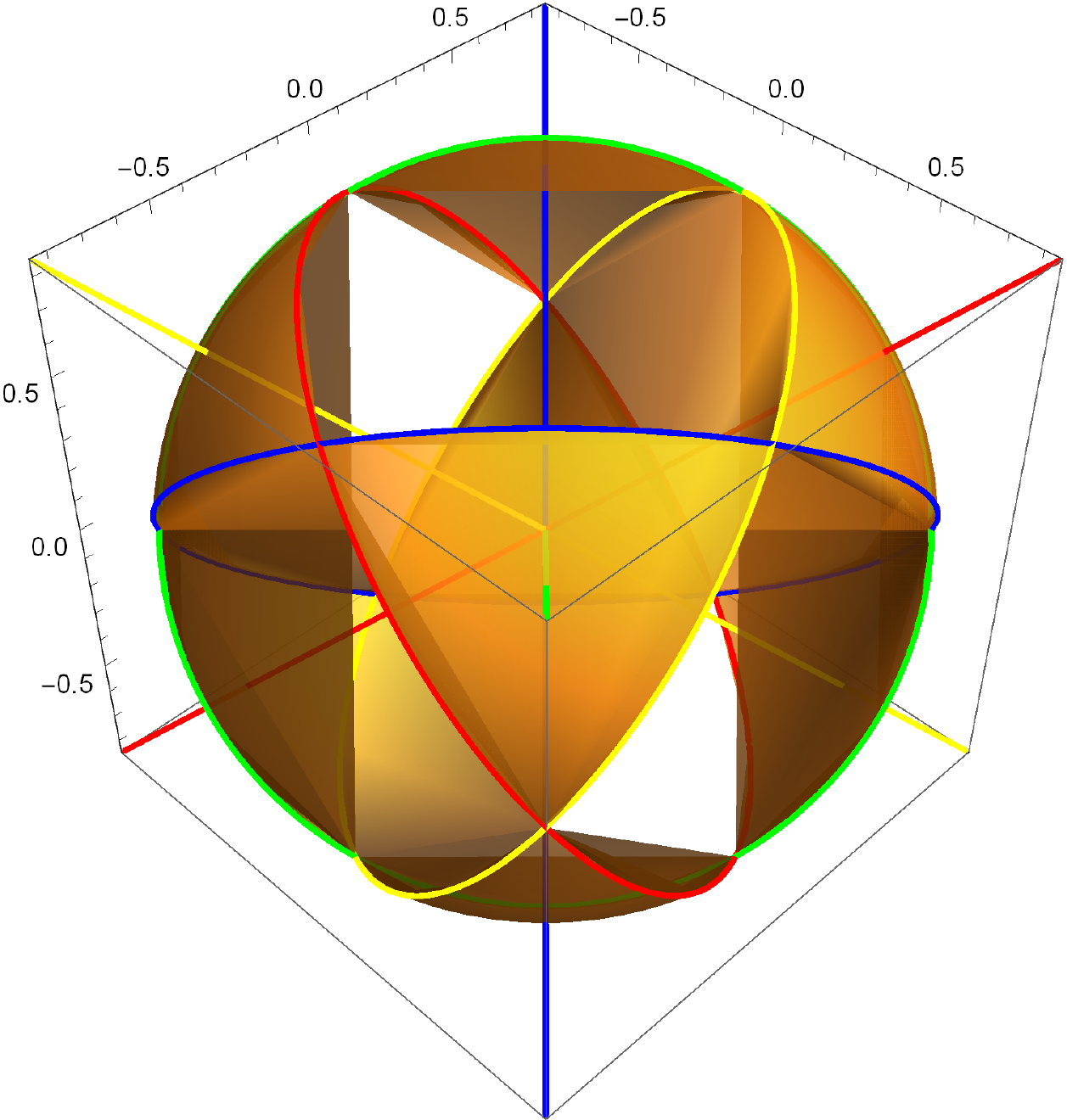}
  \caption[The projective cuboctahedron]
   {The set of normal vectors $CEN$ of completely elliptic two-dimensional subspaces of ${\mathbbm R}^3$.
   A completely elliptic subspace hosts two-dimensional spin configurations.
   In this case $CEN$ is the union of the open interior of the triangles bounded by the four (colored) great circles perpendicular
   to the four (colored) space diagonals and the set of its vertices.
   }
  \label{FIGES3}
\end{figure}

Two-dimensional subspaces of ${\mathbbm R}^3$ are conveniently represented by their normal vectors which are unique up to a sign.
Hence they should rather be considered as elements of the projective plane ${\mathbbm P}^2({\mathbbm R})$.
Let $CEN\subset {\mathbbm P}^2({\mathbbm R})$ denote the set of normal vectors of CES.
Any two-dimensional subspace ${ S}$ containing one of the lines $L_\mu$ is an elliptic subspace, although not a completely one. But it can be
obtained as a limit of CES such that the ellipses spanned by the rows of the corresponding $W$  degenerate into a pair of parallel lines.
It follows that the normal vectors of such two-dimensional subspaces ${ S}$ run through the great circle $C_\mu$ of the unit sphere
perpendicular to $L_\mu$. Hence the four great circles $C_\mu,\mu=0,\ldots,3$ form the boundary of the set $CEN$ of normals of CES.
These great circles divide the unit sphere into $8$ triangles and $6$ squares and hence define a structure that could be called
the ``projective cuboctahedron", see Figure \ref{FIGES3}. Obviously, the partial reflections and permutations operate as symmetries of the projective cuboctahedron.
It remains to decide whether the $CEN$ is generated by the open interior of the triangles or of the squares.  To this end we consider the matrix
\begin{equation}\label{ES31}
W=\left(
\begin{array}{cc}
 -\frac{1}{2} & \frac{\sqrt{3}}{2} \\
 -\frac{1}{2} & -\frac{\sqrt{3}}{2} \\
 1 & 0 \\
\end{array}
\right)\;,
\end{equation}
the columns of which span a CES with normal vector ${\mathbf w}=(\frac{1}{\sqrt{3}},\frac{1}{\sqrt{3}},\frac{1}{\sqrt{3}})$.
Since ${\mathbf w}$ lies inside the triangle of the projective cuboctahedron bounded by $C_1,\,C_2,\,C_3$  we conclude that
$CEN$ is generated by the four open interiors of the triangles (there are eight triangles but they have to be pair-wise identified
due to the undetermined sign of the normal vectors), see  Figure \ref{FIGES3}. Contrary to the first sight $CEN$ is not an open subset of
${\mathbbm P}^2({\mathbbm R})$ since the vertices of the triangles also belong to it.

It will be instructive to re-derive the above characterization of the CES of ${\mathbbm R}^3$ by means of the Lagrange variety approach.
Let $S\subset{\mathbbm R}^3$ be CES of dimension $2$ and ${\mathbf w}$ be a unit vector normal to $S$. Further, let $P$ be the projector onto
${\mathbf w}$, i.~e., $P_{i\,j}=w_i\,w_j$. Since, by construction, the support of $P$ is orthogonal to $S$, we may set
$P={\mathbbm J}(\check{\boldsymbol\kappa})$ such that the Gram matrix $G$ of a two-dimensional ground state ${\mathbf s}$ of ${\mathbbm J}$ lives on $S$, see Theorem \ref{Theorem1}, and $\check{\boldsymbol\kappa}$ are the Lagrange parameters w.~r.~t.~the ground state ${\mathbf s}$.
We expand the determinant ${\mathcal D}=\det {\mathbbm J}({\boldsymbol\kappa})$ up to second order w.~r.~t.~the parameters
${\mathbf x}\equiv {\boldsymbol\kappa}-\check{\boldsymbol\kappa}$ and obtain
\begin{equation}\label{ES32}
 {\mathcal D}=\frac{1}{2}\langle {\mathbf x}\left| M\right| {\mathbf x}\rangle +O(||x||^3)
 \;,
\end{equation}
where
\begin{equation}\label{ES33}
M=\left(
\begin{array}{ccc}
 0 & w_3^2 & w_2^2 \\
 w_3^2 & 0 & w_1^2 \\
 w_2^2 & w_1^2 & 0 \\
\end{array}
\right)\;.
\end{equation}

In the generic case, that is for $\det M\neq 0$,  $\langle {\mathbf x}\left| M\right| {\mathbf x}\rangle =0$
defines an elliptic double cone ${\mathcal C}$ that approximates
the variety ${\mathcal D}=0$ in the neighborhood of ${\mathbf x}={\mathbf 0}$. For $\det M =0$ this cone
degenerates into a wedge spanned by two planes, see Figure \ref{FST5} for an example. According to Theorem \ref{Theorem1},\, $S$ is elliptic iff the cone
${\mathcal C}$ is vertical. An equivalent condition for verticality is that the plane ${\mathbf e}^\perp$ perpendicular to ${\mathbf e}=(1,1,1)$
supports ${\mathcal C}$, which is satisfied if ${\mathbf e}^\perp$ intersects ${\mathcal C}$ only at the origin. Hence we look for solutions of
\begin{eqnarray}\label{ES34a}
 0&=&\left(
\begin{array}{c}
x_1  \\
x_2  \\
 -x_1-x_2 \\
\end{array}
\right)
\cdot
M
\cdot
\left(
\begin{array}{c}
x_1  \\
x_2  \\
 -x_1-x_2 \\
\end{array}
\right) \\
\nonumber
&=&
\scriptsize
\left(
\begin{array}{c}
x_1  \\
x_2  \\
\end{array}
\right)
\cdot
\left(
\begin{array}{cc}
 -2 w_2^2 & -w_1^2-w_2^2+w_3^2 \\
 -w_1^2-w_2^2+w_3^2 & -2 w_1^2 \\
\end{array}
\right)
\cdot
\left(
\begin{array}{c}
x_1  \\
x_2  \\
\end{array}
\right)\\
\label{ES34b}
&&\\
\label{ES34c}
&\equiv&
\left(
\begin{array}{c}
x_1  \\
x_2  \\
\end{array}
\right)
\cdot
A\cdot
\left(
\begin{array}{c}
x_1  \\
x_2  \\
\end{array}
\right)
\;.
\end{eqnarray}
This equation has only the trivial solution $x_1=x_2=0$ iff $\det A>0$. After some calculations we obtain
\begin{eqnarray}
\label{ES35a}
  \det A &=& -w_1^4+2 \left(w_2^2+w_3^2\right) w_1^2-\left(w_2^2-w_3^2\right)^2 \\
 \nonumber
   &=& (w_1+w_2-w_3) (w_1-w_2+w_3)\\
    \label{ES35b}
   && (-w_1+w_2+w_3) (w_1+w_2+w_3)
   \;.
\end{eqnarray}

Hence $\det A>0$ describes the open interior of the four triangles bounded by the great circles $C_0,\,C_1,\,C_2,\,C_3$
mentioned above and thus characterizes $CEN$ except for the vertices of the triangles. It is interesting to note that
the polynomial (\ref{ES35b}) is a symmetric polynomial not only w.~r.~t.~permutations but also w.~r.~t.~the
partial reflections  $\rho_{\mathcal T}$.

According to Theorem \ref{Theorem1} we have the bijections $\mbox{\sf Fac}({\mathcal G})\leftrightarrow{\mathbb E}\leftrightarrow CES$.
It will be interesting to identify the bijection 
$\mbox{\sf Fac}({\mathcal G})\leftrightarrow CES$
for the general spin triangle. The interior of the four triangles that together generate $CEN$
corresponds to the four connected components of the manifold $M$  defined in (\ref{FST2e}) consisting of co-planar ground states.
Each connected component of $M$ is enclosed by three line segments $\overline{A_i,A_j}$ of the form (\ref{FST2f}) -- (\ref{FST2k})
consisting of families of co-planar ground states with two spin vectors parallel or anti-parallel. Since these families generate the same
CES they correspond to the three vertices of the above-mentioned triangles, resp., that connect each triangle to its three neighbors. Finally,
as we have already seen, the great circles $C_\mu,\mu=0,\ldots,3$ correspond to Ising ground states and hence to the four extremal points
$A_i,\,i=0,1,2,3$ defined in (\ref{FST2a}). Each great circle $C_\mu$ touches three triangles of the projective cuboctahedron in the same
way as each extremal point $A_i$ touches three of the four connected components of $M$.

\section{The group of symmetries of the Gram set}\label{sec:SY}

Recall that the Gram set ${\mathcal G}={\mathcal G}_N$ is an ${N\choose 2}$-dimensional compact convex set. The group of affine bijections,
or ``symmetries",
$\sigma: {\mathcal G}\longrightarrow{\mathcal G}$ will preliminary be denoted by ${\sf Aut}$. We have shown in section \ref{sec:DGS}
that permutations $\pi$ of $N$ spins and partial reflections $\rho_{\mathcal T},\;{\mathcal T}\subset\{1,\ldots,N\},\,$
generate affine bijections of ${\mathcal G}$ via
$G\mapsto \Pi(G)= U(\pi)^{-1}\,G\,U(\pi)$ and $G\mapsto {\sf R}_{\mathcal T}(G)=V(\rho_{\mathcal T})\,G\,V(\rho_{\mathcal T})$.
These special symmetries form subgroups ${\sf Per}$ and ${\sf Rf}$ of ${\sf Aut}$. The
finite subgroup of ${\sf Aut}$ generated by ${\sf Per}$ and ${\sf Rf}$ has been denoted by ${\sf Sym}$. We note the following:
\begin{lemma}\label{lemmaN}
 ${\sf Rf}$ is a normal subgroup of ${\sf Sym}$.
\end{lemma}

\noindent{\bf Proof}\\
It suffices to show that for all $\Pi\in{\sf Per}$ and ${\mathcal T}\subset \{1,\ldots,N\}$ there exists a
${\mathcal T}'\subset \{1,\ldots,N\}$  such that
 $\Pi\,{\sf R}_{\mathcal T}\,\Pi^{-1}={\sf R}_{{\mathcal T}'}$. This holds for
 ${\mathcal T}'\equiv \{\pi(\mu)\left| \mu\in{\mathcal T}\right.\}$ which implies
 $V(\rho_{{\mathcal T}'})=U_\pi^{-1}\,V(\rho_{\mathcal T})\,U_\pi$ and further
 \begin{eqnarray}\nonumber
 \Pi\,{\sf R}_{\mathcal T}\,\Pi^{-1}(G)&=&U_\pi^{-1}\,V({\rho_{\mathcal T}})\, U_\pi\,G\,U_\pi^{-1}\,V({\rho_{\mathcal T}})\, U_\pi\\
 \label{SY1b}
&=&V({\rho_{{\mathcal T}'}})\,G\,V({\rho_{{\mathcal T}'}})={\sf R}_{{\mathcal T}'}(G)
\;.
 \end{eqnarray}

 \hfill$\Box$\\

As the main result of this section we will prove that permutations and partial reflections already generate the group of all symmetries:
\begin{theorem}\label{TheoremSY}
 ${\sf Aut}$ = ${\sf Sym}$.
\end{theorem}

The proof of this theorem will be split into a couple of lemmas.

\begin{lemma}\label{lemmaSY1}
Every symmetry $\sigma\in {\sf Aut}$ maps ${\mathcal I}$ onto ${\mathcal I}$.
\end{lemma}

\noindent{\bf Proof}\\
It is clear that  $\sigma\in {\sf Aut}$ maps extremal points of ${\mathcal G}$ onto extremal points.
Ising matrices are exactly the extremal points $G_0$ of ${\mathcal G}$ with $\mbox{rank }G_0=1$ and
hence the closed faces of the form $\{G_0\}={\sf fac}(S)$ where the completely elliptic subspace $S$
is one-dimensional and thus $S^\perp$ is $(N-1)$-dimensional.
Then the property $\mbox{rank }G_0=1$ can be cast into the language of affine geometry
by the equivalent statement that the cone ${\mathcal C}(G_0)$ defined in Theorem \ref{Theorem1a} (iii)
has the maximal dimension $\frac{1}{2}N(N-1)$.
Consequently this property is respected by symmetries $\sigma\in {\sf Aut}$
that hence generate permutations of ${\mathcal I}$.                       \hfill$\Box$\\

Lemma \ref{lemmaSY1} immediately implies:
\begin{lemma}\label{lemmaSY2}
The barycenter ${\mathbbm 1}_N$ of ${\mathcal I}$ is a fixed point for all $\sigma\in {\sf Aut}$.
\end{lemma}

It will be convenient to replace  ${\mathcal G}$ by the affinely isomorphic convex set ${\mathcal G}-{\mathbbm 1}_N$.
Consequently, the symmetries $\sigma\in {\sf Aut}$ leave
the point ${\mathbbm O}$, the barycenter of ${\mathcal I}$,
fixed and hence can be considered as linear transformations of the linear space ${\mathcal L}$ spanned by ${\mathcal G}$.
As a consequence in the remainder of this section we will usually make no difference between, say, $A$ and $\hat{A}=A-{\mathbbm 1}$.

\begin{lemma}\label{lemmaSYL}
The linear span of ${\mathcal I}$ is ${\mathcal L}$, the linear space of
all symmetric $N\times N$-matrices with vanishing diagonal and hence of dimension ${N\choose 2}$.
\end{lemma}

\noindent{\bf Proof}\\
We will prove this lemma by induction over $N$. For $N=2$ the two matrices
$\hat{A_0}=\left(\begin{array}{cc}
0&1\\
1&0
\end{array}
\right)$
and
$\hat{A_0}=\left(\begin{array}{cc}
0&-1\\
-1&0
\end{array}
\right)$
are linearly dependent and $\dim {\mathcal L}=1$.

For the induction step $N\rightarrow N+1$ we may assume that the linear span of ${\mathcal I}_N$
is  ${\mathcal L}_N$ and $\dim {\mathcal L}_N={N\choose 2}$.
The notation ${\mathcal I}_N$ denoting the set of Ising matrices for $N$ spins  must not be confused with the
notation ${\mathcal I}_k$ denoting sets of $k$-Ising matrices introduced in section \ref{sec:DGSI}.
For each Ising state ${\mathbf s}$ of
$N$ spins we obtain two Ising states ${\mathbf s}+$ and ${\mathbf s}-$ of $N+1$ spins by choosing ${\mathbf s}_{N+1}=+1$ or $-1$, resp.~.
For the Ising matrices $\hat{G}({\mathbf s}+)$ and $\hat{G}({\mathbf s}-)$ this means that $\hat{G}({\mathbf s})$ is augmented
by a copy of its first row (column), or by a negative copy of its first row (column), except the marginal values
$\hat{G}({\mathbf s}\pm)_{1,N+1}=\hat{G}({\mathbf s}\pm)_{N+1,1}=\pm 1$ and $\hat{G}({\mathbf s}\pm)_{N+1,N+1}=0$.
Let ${\mathbf s}$ run through a set $I_N$  of ${N\choose 2}$ Ising states such that the corresponding set
$\{\hat{G}({\mathbf s})\left| {\mathbf s}\in I_N\right.\}$ forms a basis of ${\mathcal L}_N$.
Then the set ${\mathcal B}\equiv\{\frac{1}{2}\left( \hat{G}({\mathbf s}+)+\hat{G}({\mathbf s}-)\right)\left.\right|{\mathbf s}\in I_N\}$ consists of
copies of the matrices in the above basis of ${\mathcal L}_N$ augmented by zero rows/columns. Hence it is a linearly independent set in
${\mathcal L}_{N+1}$. Recall that the sum of all matrices in ${\mathcal I}_N$ yields the zero matrix ${\mathbbm O}$.
It follows that the corresponding sum of $(N+1)\times (N+1)$-matrices $\hat{G}({\mathbf s}+)$, multiplied with $\frac{1}{|{\mathcal I}_N|}$,
yields a matrix $H=H^{(1)}$ with zero entries except $H_{N+1,1}=H_{1,N+1}=1$. It follows that $H$ cannot be obtained as a linear combination
of matrices from ${\mathcal B}$, hence ${\mathcal B}\cup\{H\}$ is still linearly independent. Let $\pi=(1,\mu)$, the transposition swapping
$1$ and $\mu,\;2\le \mu\le N$. The corresponding symmetry $H\mapsto \Pi(H)$ swaps the first and the $\mu$-th row/column of $G$.
Hence the $N-1$ resulting matrices $H^{(\mu)}\equiv \Pi(H)$, that are in the linear span of ${\mathcal I}_{N+1}$, have only zero entries
except $\Pi(H)_{N+1,\mu}=H_{\mu,N+1}=1$. The set ${\mathcal A}\equiv\{H^{(\mu)}| \mu=1,\ldots,N\}$ is hence linearly independent
in ${\mathcal L}_{N+1}$ and  ${\mathcal A}\cup {\mathcal B}$ forms a basis of ${\mathcal L}_{N+1}$ with ${N\choose 2}+N={N+1\choose 2}$
elements.  \hfill$\Box$\\

\begin{lemma}\label{lemmaSY4}
If $\sigma\in {\sf Aut}$ leaves all points of ${\mathcal I}$ fixed then it is the identity.
\end{lemma}

\noindent{\bf Proof}\\
This follows immediately from Lemma \ref{lemmaSYL} since $\sigma$ is a linear map and under the assumption of the lemma it leaves all points of ${\mathcal L}$, the linear span of ${\mathcal I}$, fixed.      \hfill$\Box$\\

It follows from Lemma \ref{lemmaSY4} that a symmetry $\sigma\in {\sf Aut}$ is completely determined by its action on
${\mathcal I}$. Indeed, for every other symmetry $\tau\in {\sf Aut}$ that coincides with $\sigma$ on ${\mathcal I}$,
it follows that $\tau^{-1}\,\sigma$ leaves all points of ${\mathcal I}$ fixed, and, by Lemma \ref{lemmaSY4}, $\sigma=\tau$.
This further implies that ${\sf Aut}$ is isomorphic to a subgroup of the group of permutations of ${\mathcal I}$
and hence finite. Hence we have shown:
\begin{lemma}\label{lemmaSY5}
${\sf Aut}$ is a finite group.
Every symmetry $\sigma\in {\sf Aut}$ is completely determined by its action on ${\mathcal I}$.
\end{lemma}

One easily proves that the subgroup ${\sf Sym}$ of ${\sf Aut}$ consists of
orthogonal transformations w.~r.~t.~the scalar product $\mbox{Tr }A\,B$,
but the same property need not hold for the possibly larger group ${\sf Aut}$. For this reason we define another
Euclidean scalar product
$\prec\,|\,\succ$
on the linear space  ${\mathcal L}$ by
averaging the former scalar product over the action of ${\sf Aut}$:
\begin{equation}\label{SY3}
\prec A |B\succ\equiv \frac{1}{|{\sf Aut}|}\sum_{\sigma\in{\sf Aut}}\mbox{Tr }(\sigma (\hat{A})\,\sigma (\hat{B}))
\;.
\end{equation}
At the r.~h.~s.~of (\ref{SY3}) we have made use of the matrix properties of the points of ${\mathcal L}$
and consequently re-introduced the $\hat{}\;$-notation.
It follows that all symmetries $\sigma\in {\sf Aut}$ are $\prec \,|\,\succ$-orthogonal
transformations,
whereas the $\sigma\in {\sf Sym}$ are additionally orthogonal transformations w.~r.~t.~the former scalar product $\mbox{Tr }A\,B$.
Nevertheless, the following holds:
\begin{lemma}\label{lemmaSY6}
All distances between Ising matrices calculated in section \ref{sec:DGSI}
have the same value also if calculated w.~r.~t.~the scalar product $\prec \,|\,\succ$ defined in (\ref{SY3}).
\end{lemma}

\noindent{\bf Proof}\\
First we show that the norm of Ising matrices is the same for both scalar products. This follows since
$\sigma ({\boldsymbol\gamma})\in{\mathcal I}$ for all ${\boldsymbol\gamma}\in{\mathcal I}$ and hence
$\mbox{Tr }(\sigma (\hat{\boldsymbol\gamma})\,\sigma (\hat{\boldsymbol\gamma}))=N(N-1)$ for all $\sigma\in{\sf Aut}$.
Therefore the averaging in (\ref{SY3}) of the constant value yields
$\prec\hat{\boldsymbol{\gamma}} |\hat{\boldsymbol\gamma}\succ=N(N-1)$.

Next we consider the representation of the barycenter $B_k$ of ${\mathcal I}_k$, see (\ref{DGSI4}),
that can be re-written as $B_k=\alpha A_0$. This representation is independent of the choice of the scalar product,
but it is not clear whether $B_k$ is the\\ $\prec \,|\,\succ$-projection of all
${\boldsymbol\gamma}\in{\mathcal I}_k$ onto the line through $A_0$. Hence let $C_k=\beta A_0$ be
the $\prec \,|\,\succ$-projection of an arbitrary
${\boldsymbol\gamma}\in{\mathcal I}_k$ onto the line through $A_0$ and let ${\boldsymbol\gamma}'\in{\mathcal I}_k$
be a different point. Then there exists a permutation $\pi\in S_N$ such that the corresponding symmetry $\Pi$
maps ${\boldsymbol\gamma}$ onto ${\boldsymbol\gamma}'$ and leaves $A_0$ fixed. Since $\Pi$ is also a
$\prec \,|\,\succ$-orthogonal transformation, ${\boldsymbol\gamma}'$ has the
same $\prec \,|\,\succ$-projection onto the line through $A_0$ as ${\boldsymbol\gamma}$.
Consequently all ${\boldsymbol\gamma}\in{\mathcal I}_k$ and hence also the barycenter $B_k$ have the same projection
$C_k=\beta A_0$ onto the line through $A_0$. But $B_k$ is already of the form $B_k=\alpha A_0$ hence $B_k=C_k$
and we have shown that the projection does not depend on the scalar product.

From this it follows immediately that (\ref{DGSI2}) and  (\ref{DGSI3}) also hold for the new scalar product
$\prec \,|\,\succ$. For the distances between arbitrary pairs of points
in ${\mathcal I}_k$ the analogous statement follows by application of the symmetries ${\sf R}_{\mathcal T}$ generated by partial reflections
$\rho_{\mathcal T}$ that are isometries w.~r.~t.~either scalar product.       \hfill$\Box$\\

\begin{lemma}\label{lemmaSY7}
If a symmetry $\sigma\in{\sf Aut}$ leaves $A_0$ fixed then it maps the sets ${\mathcal I}_k$ onto  ${\mathcal I}_k$
for all $k=1,\ldots,\lfloor \frac{N}{2}\rfloor$.
\end{lemma}

\noindent{\bf Proof}\\
The lemma follows since, by (\ref{DGSI3}), the sets ${\mathcal I}_k$ are exactly the subsets of ${\mathcal I}$ that have
a constant distance $2\,\sqrt{2\, k\,(N-k)}$ to $A_0$ and since $\sigma\in{\sf Aut}$ is
an $\prec \,|\,\succ$-isometry. Note that also Lemma \ref{lemmaSY6} is needed
for insuring that the above distances are the same for both scalar products. \hfill$\Box$\\

\begin{lemma}\label{lemmaSY8}
If a symmetry $\sigma\in{\sf Aut}$ leaves $A_0$ and all $A_\mu\in{\mathcal I}_1,\;\mu=1,\ldots,N,$ fixed then it is the identity.
\end{lemma}

\noindent{\bf Proof}\\
By Lemma \ref{lemmaSY7} the sets ${\mathcal I}_k$ are left invariant by $\sigma$.
It suffices to show that $\sigma$ is the identity for all  ${\mathcal I}_k,\;k=2,\ldots,\lfloor \frac{N}{2}\rfloor$
since it is completely determined by its action on ${\mathcal I}$, see Lemma \ref{lemmaSY5}.
We will prove the claim by induction over $k$ and start with $k=2$. Since we will make use of the matrix
properties of the points of  ${\mathcal I}$ it will be appropriate to distinguish between, say, $A$ and $\hat{A}=A-{\mathbbm 1}$
and to undo the translation ${\mathcal G}\mapsto {\mathcal G}-{\mathbbm 1}$
for the remainder of this proof.

Let $A_{\{\mu,\nu\}},\; 1\le \mu<\nu\le N$ be an arbitrary element of ${\mathcal I}_2$. Recall that $A_{\{\mu,\nu\}}$
is the Gram matrix of an Ising spin configuration ${\mathbf s}$ where, say,  all spins are $\uparrow$ except
${\mathbf s}_\mu={\mathbf s}_\nu=\downarrow$. Consider the partial reflection $\rho_{\{\mu\}}$ and the corresponding symmetry
${\sf R}_{\{\mu\}}\in {\sf Sym}$. Since $\rho_{\{\mu\}}$ inverts the spin ${\mathbf s}_\mu$ we have
${\sf R}_{\{\mu\}} (A_{\nu})=A_{\{\mu,\nu\}}$, analogously ${\sf R}_{\{\nu\}} (A_{\mu})=A_{\{\mu,\nu\}}$.
It follows that
\begin{equation}\label{SY4}
 A_{\{\mu,\nu\}}\in {\sf R}_{\{\mu\}}\left[ {\mathcal I}_1\right] \cap {\sf R}_{\{\nu\}}\left[ {\mathcal I}_1\right]
 \;.
\end{equation}
${\mathcal I}_1$ is the set of points ${\boldsymbol\gamma}\in {\mathcal I}$ that have the constant distance
$ 2\,\sqrt{2\, (N-1)}$ from $A_0$. Since ${\sf R}_{\{\mu\}}$ is an isometry we may characterize
${\sf R}_{\{\mu\}}\left[ {\mathcal I}_1\right] $ as the set of points ${\boldsymbol\gamma}\in {\mathcal I}$ that have the constant distance
$ 2\,\sqrt{2\, (N-1)}$ from ${\sf R}_{\{\mu\}}(A_0)=A_\mu$. By assumption, $\sigma$ leaves $A_\mu$ fixed and, being an isometry,
maps ${\sf R}_{\{\mu\}}\left[ {\mathcal I}_1\right] $ onto itself.
Analogously, $\sigma$ maps ${\sf R}_{\{\nu\}}\left[ {\mathcal I}_1\right] $ onto itself. Applying $\sigma$ to (\ref{SY4}) hence yields
\begin{equation}\label{SY5}
\sigma\left( A_{\{\mu,\nu\}}\right)\in {\sf R}_{\{\mu\}}\left[ {\mathcal I}_1\right] \cap {\sf R}_{\{\nu\}}\left[ {\mathcal I}_1\right]
 \;.
\end{equation}
We will finish the initial step of the induction over $k$ by showing that
${\sf R}_{\{\mu\}}\left[ {\mathcal I}_1\right] \cap {\sf R}_{\{\nu\}}\left[ {\mathcal I}_1\right]$
is a set with exactly two elements, $A_{\{\mu,\nu\}}$ and $A_0$, the latter being fixed under $\sigma$. Then (\ref{SY5}) would imply
that $A_{\{\mu,\nu\}}$ will be a fixed point of $\sigma$.

To this end we argue that an arbitrary element of ${\sf R}_{\{\mu\}}\left[ {\mathcal I}_1\right]$ is either of the form
$A_{\{\mu,\lambda\}},\;1\le \lambda\le N$ and $\lambda\neq \mu$ or of the form $A_0$. Analogously, the general element of ${\sf R}_{\{\nu\}}\left[ {\mathcal I}_1\right]$
is of the form $A_{\{\nu,\kappa\}},\;1\le \kappa\le N$ and $\kappa\neq \nu$ or of the form $A_0$. Hence any element ${\boldsymbol\gamma}$ in the intersection
${\boldsymbol\gamma}\in{\sf R}_{\{\mu\}}\left[ {\mathcal I}_1\right] \cap {\sf R}_{\{\nu\}}\left[ {\mathcal I}_1\right]$
is equal to $A_0$ or satisfies
${\boldsymbol\gamma}=A_{\{\mu,\lambda\}}=A_{\{\nu,\kappa\}}$ which is only possible if $\lambda=\nu$ and $\kappa=\mu$, i.~e.,
${\boldsymbol\gamma}=A_{\{\mu,\nu\}}$. This completes the initial step of the proof by induction over $k$.

For the step $k\rightarrow k+1$ we will assume that all points of  ${\mathcal I}_n$ are fixed points of $\sigma$ for all $n=1,\ldots,k$.
Let $A_{\mathcal K}$ be an arbitrary element of ${\mathcal I}_{k+1}$.
Recall that $A_{\mathcal K},\;{\mathcal K}\subset\{1,\ldots,N\}$
is the Gram matrix of an Ising spin configuration ${\mathbf s}$ where, say,  all spins are $\uparrow$ except
${\mathbf s}_\mu=\downarrow$ for $\mu\in{\mathcal K}$.

Choose two indices $\mu,\nu\in{\mathcal K},\;\mu\neq\nu$
that will be fixed for the remaining proof and define the subsets ${\mathcal K}_\mu,\;{\mathcal K}_\nu$ by removing
$\mu$, resp.~$\nu$, from ${\mathcal K}$, i.~e., ${\mathcal K}_\mu\equiv {\mathcal K}\backslash\{\mu \}$
and
${\mathcal K}_\nu\equiv {\mathcal K}\backslash\{\nu \}$ ,
such that
${\mathcal K}={\mathcal K}_\mu\cup\{\mu\}={\mathcal K}_\nu\cup\{\nu\}$.

Consider the partial reflection $\rho_{{\mathcal K}_\mu}$ and the corresponding symmetry
${\sf R}_{{\mathcal K}_\mu}\in {\sf Sym}$. Since $\rho_{{\mathcal K}_\mu}$ inverts the spins ${\mathbf s}_\lambda,\;\lambda\in{\mathcal K}_\mu$ we have
${\sf R}_{{\mathcal K}_\mu} (A_{\mu})=A_{\mathcal K}$, analogously ${\sf R}_{{\mathcal K}_\nu} (A_{\nu})=A_{\mathcal K}$.
It follows that
\begin{equation}\label{SY6}
A_{\mathcal K}\in {\sf R}_{{\mathcal K}_\mu}\left[ {\mathcal I}_1\right] \cap {\sf R}_{{\mathcal K}_\nu}\left[ {\mathcal I}_1\right]
 \;.
\end{equation}
${\mathcal I}_1$ is the set of points ${\boldsymbol\gamma}\in {\mathcal I}$ that have the constant distance
$ 2\,\sqrt{2\, (N-1)}$ from $A_0$. Since ${\sf R}_{{\mathcal K}_\mu}$ is an isometry we may characterize
${\sf R}_{{\mathcal K}_\mu}\left[ {\mathcal I}_1\right] $ as the set of points ${\boldsymbol\gamma}\in {\mathcal I}$ that have the constant distance
$ 2\,\sqrt{2\, (N-1)}$ from ${\sf R}_{{\mathcal K}_\mu}(A_0)=A_{{\mathcal K}_\mu}$.
By induction assumption, $\sigma$ leaves $A_{{\mathcal K}_\mu}$ fixed and, being an isometry,
maps ${\sf R}_{{\mathcal K}_\mu}\left[ {\mathcal I}_1\right] $ onto itself.
Analogously, $\sigma$ maps ${\sf R}_{{\mathcal K}_\nu}\left[ {\mathcal I}_1\right] $ onto itself. Applying $\sigma$ to (\ref{SY6}) hence yields
\begin{equation}\label{SY7}
\sigma\left( A_{\mathcal K}\right)\in {\sf R}_{{\mathcal K}_\mu}\left[ {\mathcal I}_1\right] \cap {\sf R}_{{\mathcal K}_\nu}\left[ {\mathcal I}_1\right]
 \;.
\end{equation}
We will finish the induction step by showing that
${\sf R}_{{\mathcal K}_\mu}\left[ {\mathcal I}_1\right] \cap {\sf R}_{{\mathcal K}_\nu}\left[ {\mathcal I}_1\right]$
is a set with exactly two elements, $A_{\mathcal K}$ and some $A_{{\mathcal K}_0}\in{\mathcal I}_{k-1}$, the latter being fixed under $\sigma$
by induction assumption.
Then (\ref{SY7}) would imply that $A_{\mathcal K}$ will be a fixed point of $\sigma$.

To this end we argue that the general element ${\boldsymbol\gamma}$ of ${\sf R}_{{\mathcal K}_\mu}\left[ {\mathcal I}_1\right] $ is of the form
${\boldsymbol\gamma}={\sf R}_{{\mathcal K}_\mu}\left( A_\lambda\right),\;1\le\lambda\le N$. If $\lambda\in {\mathcal K}_\mu$ then
${\boldsymbol\gamma}=A_{{\mathcal K}_0}$, where ${\mathcal K}_0\equiv{\mathcal K}_\mu \backslash \{\lambda\}$ and hence
$\left|{\mathcal K}_0\right|=k-1$. In the case $\lambda\notin {\mathcal K}_1$ we obtain
${\boldsymbol\gamma}=A_{{\mathcal K}_2}$ where ${\mathcal K}_2\equiv{\mathcal K}_\mu \cup \{\lambda\}$ and hence
$\left|{\mathcal K}_2\right|=k+1$.

 Analogously, the general element of ${\boldsymbol\gamma}$ of ${\sf R}_{{\mathcal K}_\nu}\left[ {\mathcal I}_1\right] $ is of the form
${\boldsymbol\gamma}={\sf R}_{{\mathcal K}_\nu}\left( A_\kappa\right),\;1\le\kappa\le N$. If $\kappa\in {\mathcal K}_\nu$ then
${\boldsymbol\gamma}=A_{{\mathcal K}_1}$, where ${\mathcal K}_1\equiv{\mathcal K}_\nu \backslash \{\kappa\}$ and hence
$\left|{\mathcal K}_1\right|=k-1$. In the case $\kappa\notin {\mathcal K}_\nu$ we obtain
${\boldsymbol\gamma}=A_{{\mathcal K}_3}$ where ${\mathcal K}_3\equiv{\mathcal K}_\nu \cup \{\kappa\}$ and hence
$\left|{\mathcal K}_3\right|=k+1$.

Let ${\boldsymbol\gamma}$ be an arbitrary element of the intersection
${\sf R}_{{\mathcal K}_\mu}\left[ {\mathcal I}_1\right] \cap {\sf R}_{{\mathcal K}_\nu}\left[ {\mathcal I}_1\right]$.
According to the preceding considerations we have either ${\boldsymbol\gamma}\in {\mathcal I}_{k-1}$
or ${\boldsymbol\gamma}\in {\mathcal I}_{k+1}$. In the first case we conclude ${\boldsymbol\gamma}=A_{{\mathcal K}_0}=A_{{\mathcal K}_1}$
and ${\boldsymbol\gamma}$ is fixed under $\sigma$. This is only possible for $\lambda=\nu$ and $\kappa=\mu$ and thus gives
exactly one point ${\boldsymbol\gamma}=A_{{\mathcal K}_0}$ of the intersection.

In the second case we conclude ${\boldsymbol\gamma}=A_{{\mathcal K}_2}=A_{{\mathcal K}_3}$, and hence
${\mathcal K}_2={\mathcal K}_\mu \cup \{\lambda\}={\mathcal K}_\nu \cup \{\kappa\}={\mathcal K}_3$.
This is only possible if $\lambda=\mu$ and $\kappa=\nu$ and hence ${\mathcal K}_2={\mathcal K}_3={\mathcal K}$
and thus yields the second point ${\boldsymbol\gamma}=A_{{\mathcal K}} $ of the intersection.

This completes the  proof of Lemma \ref{lemmaSY8}.  \hfill$\Box$\\

Now we can complete the proof of Theorem \ref{TheoremSY}. Let $\sigma\in{\sf Aut}$ be arbitrary.
There exists a unique symmetry
${\sf R}_{\mathcal T} \in {\sf Sym}$ induced by a partial reflection $\rho_{\mathcal T}$ such that
${\sf R}_{\mathcal T}\left(A_0\right) =\sigma\left(A_0\right)$. Hence ${\sf R}_{\mathcal T}^{-1}\,\sigma$
leaves $A_0$ fixed and hence ${\mathcal I}_1$ invariant, c.~f.~Lemma \ref{lemmaSY7}. Further there exists a unique symmetry
$\Pi \in {\sf Sym}$ induced by a permutation $\pi\in S_N$ such that
${\sf R}_{\mathcal T}^{-1}\,\sigma\left(A_\mu\right) =\Pi\left(A_\mu\right)$ for all $\mu=1,\ldots,N$.
Moreover, $\Pi^{-1}$ leaves $A_0$ fixed.
Hence $\tau\equiv\Pi^{-1}\,{\sf R}_{\mathcal T}^{-1}\,\sigma$ leaves all $A_\mu$ fixed for $\mu=0,\ldots,N$.
By lemma \ref{lemmaSY8} we conclude that $\tau$ is the identity in ${\sf Aut}$ and hence
$\sigma= {\sf R}_{\mathcal T}\,\Pi\in{\sf Sym}$. This finally proves ${\sf Aut}$=${\sf Sym}$.  \hfill$\Box$\\

In view of the Theorem \ref{TheoremSY} we will henceforward only use the notation ${\sf Sym}$ for the symmetry group of ${\mathcal G}$.
As a by-product of the proof we have proven that every symmetry $\sigma\in{\sf Sym}$ can be uniquely written as the product
$\sigma= {\sf R}_{\mathcal T}\,\Pi,$ where  ${\sf R}_{\mathcal T}\in {\sf Rf}$ and $\Pi\in {\sf Per}$. Together with
the fact that ${\sf Rf}$ is a normal subgroup of ${\sf Sym}$, see Lemma \ref{lemmaN}, this implies:
\begin{lemma}\label{lemmaSY9}
${\sf Sym}$ is the semi-direct product of ${\sf Rf}$ and ${\sf Per}$, ${\sf Sym}={\sf Rf}\rtimes {\sf Per}$.
\end{lemma}



\begin{thebibliography}{99}

\bibitem{S17}
 H.-J. Schmidt,
Theory of ground states for classical Heisenberg spin
systems I, arXiv:cond-mat1701.02489v2, (2017)

\bibitem{S17b}
 H.-J. Schmidt,
Theory of ground states for classical Heisenberg spin
systems II, arXiv:cond-mat1707.02859, (2017)

\bibitem{SL03}
 H.-J. Schmidt, M. Luban M,
Classical ground states of symmetric Heisenberg spin systems,
 J. Phys. A {\bf 36}, 6351 -- 6378 (2003)






\bibitem{BZ06}
I. Bengtsson and K. Zyczkowski,
{\em Geometry of quantum states},
Cambridge Univ. Press, Cambridge, 2006



\bibitem{L83}
G. Ludwig,
{\em Foundations of Quantum Mechanics I},
Springer-Verlag, New York, 1983




\end{thebibliography}
\end{document}